\documentclass[10pt, letterpaper]{IEEEtran}
\usepackage[square,comma,sort,numbers]{natbib}
\usepackage[latin9]{inputenc}
\usepackage{graphicx,color}
\usepackage{xspace,subfigure,epsfig}
\usepackage{xspace}
\usepackage{graphicx,color}
\usepackage{times,amsfonts,amssymb,amsmath,setspace}
\usepackage{setspace}
\usepackage{comment,url,enumerate}
\usepackage{lipsum}
\usepackage{multicol}
\usepackage[algoruled,medskip,dontprintsemicolon,linesnumbered,Algorithm]{algorithm}
\usepackage[noend]{algpseudocode}
\usepackage{float}
\usepackage{stfloats,mathtools} 

\DeclareMathAlphabet\mathbfcal{OMS}{cmsy}{b}{n}

\usepackage{bbm}\usepackage{mathrsfs}\usepackage{float}
\usepackage{cases}%\usepackage{pseudocode}

\usepackage{graphicx,url,bm,subfigure,comment}
\usepackage{epsfig,textcomp}
\usepackage{times,amsfonts,amssymb,amsmath,setspace}%,cleveref}

\newcommand{\dd}{{\rm\,d}} % differential (for integrals)
\newcommand{\1}{\mathbbm 1}

\newcommand{\Rc}{\mathcal R}

\newcommand{\Ec}{\mathcal E}
\newcommand{\Ac}{\mathcal A}
\newcommand{\Bc}{\mathcal B}
\newcommand{\Zc}{\mathcal Z}
\newcommand{\Good}{\mathcal A'}
\newcommand{\Bad}{\mathcal A''}
\newcommand{\GoodB}{\mathcal B'}
\newcommand{\BadB}{\mathcal B''}

\newtheorem{teorema}{\bf Theorem}
\newtheorem{corollario}{\bf Corollary}

\newtheorem{definizione}{\bf Definition}
\newtheorem{proposizione}{\bf Proposition}

\def\ind{{\rm 1\hspace{-0.90ex}1}}

\newcommand{\btau}{\mathbf \tau}
\newcommand{\sm}{\setminus}

\newcommand{\Tc}{{\mathcal G}_{\text{T}}} 
\newcommand{\Gc}{\mathcal G}

\newcommand{\Hc}{\mathcal H}
\newcommand{\Kc}{\mathcal K}
\newcommand{\Mc}{\mathcal P}

\newcommand{\Vc}{\mathcal V}

\newcommand{\Nc}{n}

\newcommand{\Ecb}{\mathbfcal E}
\newcommand{\Vcb} {\mathbfcal V}

\newcommand{\ls}[1]
   {\dimen0=\fontdimen6\the\font
    \lineskip=#1\dimen0
    \advance\lineskip.5\fontdimen5\the\font
    \advance\lineskip-\dimen0
    \lineskiplimit=.9\lineskip
    \baselineskip=\lineskip
    \advance\baselineskip\dimen0
    \normallineskip\lineskip
    \normallineskiplimit\lineskiplimit
    \normalbaselineskip\baselineskip
    \ignorespaces
}
\makeatother

\newtheorem{definition}{Definition}[section]
\newcommand{\EE}{\mathbb{E}} % average operator
\newcommand{\PP}{\mathbb{P}} % probability

\begin{document}
\begin{sloppypar}

% \title{Graph Deanonymization:\\ How Power-law Degree Helps}
\title{De-anonymizing scale-free social networks by percolation graph matching}
%\title{De-anonymizing social networks by percolation graph matching: the impact of power-law degree}
%\title{De-anonymizing social networks by percolation graph matching: the power of power-law degree}
\author{Carla Chiasserini $\!^\ast$, Michele Garetto$\!^\dagger $ , Emilio Leonardi$\!^\ast $\\
$\!^\ast $ Dipartimento di Elettronica, Politecnico di Torino, Torino, Italy\\
$\!^\dagger $ Dipartimento di Informatica, Universit\`{a} di Torino, Torino, Italy}
%\affiliation{}

\maketitle

\global\long\def\cU{\mathcal{U}}
\global\long\def\cS{\mathcal{S}}
\global\long\def\bx{\mathbf{x}}
\global\long\def\ba{\mathbf{a}}
\global\long\def\bb{\mathbf{b}}
\global\long\def\bt{\mathbf{t}}
\global\long\def\bu{\mathbf{u}}
\global\long\def\bh{\mathbf{h}}
\global\long\def\by{\mathbf{y}}
\global\long\def\bz{\mathbf{z}}
\global\long\def\btau{\mathbf{\tau}}
%\global\long\def\dd{\partial}
\global\long\def\sm{\setminus}
\global\long\def\1{\mathbbm1}
\global\long\def\E{\mathbbm E}
\global\long\def\P{\mathbbm P}

\newcommand{\tgifeps}[3]{
\begin{figure}[htb]
\centering
\includegraphics[width=#1cm]{#2.eps}
% \vspace{-1mm}
\caption{#3\label{fig:#2}}
\vspace{-2mm}
\end{figure}
}

%\ls{0.90}

\begin{abstract}
We address the problem of social network de-anonymization when relationships between people are described 
by scale-free graphs. In particular, we propose a rigorous, asymptotic mathematical analysis
of the network de-anonymization problem while capturing the impact of power-law node degree
distribution, which is a fundamental and quite ubiquitous feature
of many complex systems such as social networks. 
% Such graphs account for some fundamental features  
% of real social networks while still allowing for a rigorous mathematical analysis of the user identification process.  
By applying bootstrap percolation and a novel graph slicing technique,  
we prove that large inhomogeneities in the node degree lead to a dramatic reduction of the 
initial set of nodes that must be known a priori (the seeds) in order to successfully identify all other users.
We characterize the size of this set when seeds are selected using different criteria, 
and we show that their number can be as small as $n^{\epsilon}$, for
any small ${\epsilon>0}$. 
Our results are validated through simulation experiments on a real social network graph. 
\end{abstract}

\section{Introduction}\label{sec:intro}

The increasing availability of always-on connectivity on affordable portable 
devices, coupled with the proliferation of services and online social platforms,
has provided  unprecedented opportunities to interact
and exchange information among people. At the same time, electronic traces
of our communications, searches and mobility patterns, specifically  
their collection and analysis by service providers and unintended third parties,
are posing serious treats to user privacy. This fact raises a number of well known 
and hotly debated issues, which have recently caused quite a stir in the media.

A distinctive feature of this trend is the uncontrolled proliferation
of different accounts/identities associated to each individual. 
Most of us have more than one mobile subscription,  
%(the total number of cell-phone subscriptions is approaching the population size of the 
%planet \cite{ericsson14}), 
more than one email address, 
%(there exist 4 billions active
%email accounts, vs 2.5 billions email users \cite{emailstat14}),   
and a plethora of accounts (even multiple) on popular platforms 
such as Facebook, Twitter, LinkedIn and so on.
A specific issue that naturally arises in this context is the identification 
of the different identities/accounts belonging to the same individual.
This problem, which has strong implications with user privacy,
is known in the scientific literature as social network de-anonymization
(or reconciliation). 
The two most frequently cited reasons why companies/organizations 
are interested in network  de-anonymization are user profiling (for targeted advertising and 
marketing research) and national security (i.e., the prevention of terrorism and other forms of 
criminal activity).

It is fundamental to notice that privacy concerns related to de-anonymization
are very subjective: some people do not care at all about providing 
``personally identifiable information" in their service registrations,
explicitly linking their accounts \lq\lq for-free". 
%\footnote{Many  systems and applications strongly encourage (or even
%enforce) users to provide information such as email
%address/cell-phone number as part of account creation and
%maintenance, or for the purpose of backup and synchronization with
%cloud services.}. 
As we will see, such users play a fundamental role
in the network  de-anonymization problem, acting as ``seeds'' to
identify other users. 
On the other extreme, some people are totally obsessed by the idea of 
Big Brother spying into their life and compiling tons of 
%personal 
information on all of us. Such users try to hide themselves behind anonymous
identities containing the minimum possible amount of personal data
and linkage information with other identities. In the worst case (for the entity
trying to solve the  de-anonymization problem), an identity consists just of a random
identifier (e.g., a code or a label).
%\footnote{In our work, we will assume that each random identifier
%is at least guaranteed to be uniquely and statically assigned to a single user
%of a system/application}.

One recent, dramatic discovery in the network security field \cite{Narayanan}
is the following: user privacy (in terms of anonymity) 
cannot be guaranteed by just resorting to anonymous identifiers.
In particular, the identities used by a user across different systems can be
matched together by using only the network structure
of the communications made by users (i.e., electronic traces of who
has come in contact with whom).
More formally, considering just the simple case of two systems, the 
(disordered) vertices of two social network graphs $G_1$ and $G_2$, whose edges represent
the observed contacts among users in the two systems, can be 
perfectly matched under very mild conditions on the graph structures \cite{pedarsani}.

As already anticipated, the complexity of the network  de-anonymization problem can 
be greatly reduced by having an initial (even small) number
of users already correctly matched (the seeds). Such initial side information is often indeed available, 
thanks to users who have explicitly linked their accounts, to the presence of compromised or 
fake users, as well as other forms of external information providing total or partial 
correlations among identities. Starting from the seeds, one can design
clever algorithms to progressively expand the set of matched vertices, incurring
only negligible probability to match wrong pairs \cite{lattanzi}.
 
In previous work \cite{Grossglauser}, the number of seeds
that allow to de-anonymize two networks has been 
characterized for the case of Erd\"{o}s--R\'{e}nyi random graphs,
adopting a convenient probabilistic model for 
$\Gc_1$ and $\Gc_2$. 
%from the (unknown) graph $\Gc$ describing the  complete set of human relationships among people 
%(more details later).
By reducing the graph matching problem to
a bootstrap percolation problem, authors identify a phase transition
in the number of seeds required by their algorithm.
In particular, in the case of sparse networks with average vertex degree $\Theta(\log n)$, 
the number of seeds that are provably sufficient to match the vertices scales as
$\Theta(\frac{n}{\log^{4/3} n})$, which is only a poly-log factor less than $n$.
One obvious limitation of the results in \cite{Grossglauser}
is that they apply only to Erd\"{o}s--R\'{e}nyi random graphs,
which are a poor representation of real social networks.

{\bf Contribution}. In our work we  extend the results in \cite{Grossglauser}
by considering a family of random graphs that incorporates one of the most
fundamental properties of real social networks (and many other complex systems) not yet
considered in analytical work, namely, the scale-free vertex degree distribution \cite{barabasi}.  
%Furthermore we wish remark that differently from \cite{Grossglauser}  the 
%average vertex degree in our graphs is finite.

We propose a novel algorithm for graph matching, hereinafter referred to as degree-driven 
graph matching (DDM), and show that DDM 
successfully matches a large fraction of the nodes.
% all relevant vertices, i.e., vertices with sufficiently large degree. 
%its effectiveness for the above class of random graphs.  
Similarly to \cite{Grossglauser}, we are interested in the scaling 
law of the number of seeds that are needed to make
the nodes' identification process \lq percolate', i.e., 
to propagate almost to the entire set of nodes.  

Our results mark a striking difference with those obtained for Erd\"{o}s--R\'{e}nyi graphs.
In particular, when initial seeds are uniformly distributed
among the vertices, order of $n^{\frac{1}{2}+\epsilon}$  seeds (for an  an arbitrarily small $\epsilon$)
 are sufficient to match most of the vertices.  Even more amazing results hold when initial seeds can be 
chosen (e.g., by the attacker) considering their degree: in this case,
as few as $n^\epsilon$ seeds are sufficient. The implications of our results are clear: scale-free social networks
can be surprisingly simple to match (i.e., de-anonymize), especially
when initial seeds are properly selected among the population.

Moreover, scale-free networks appear to be so amenable to  de-anonymization
that, differently from \cite{Grossglauser}, we can establish 
our results even in the case of finite average node degree (i.e., 
we do not need any densification assumption, which is necessary   
in Erd\"{o}s--R\'{e}nyi graphs if only to guarantee connectivity).
We remark that an algorithm to match scale-free networks has
been recently proposed in \cite{lattanzi}. However, in \cite{lattanzi}
authors 
%do not compute scaling laws for the minimum required number
%of seeds, because they 
do not identify any phase transition effect 
related to bootstrap percolation. Actually, they consider a 
simple direct identification strategy that requires $\Omega(\frac{n}{\log n})$ seeds
and essentially prove that their algorithm
is unlikely to match wrong pairs. Also, their analysis
is  complicated by the adoption of the preferential attachment
model by Barab\'{a}si and Albert \cite{barabasi}, whereas 
here we adopt a different model of scale-free networks
that greatly simplifies the analysis.

Finally, we emphasize that our model captures, in isolation, only
the impact of power-law degree, without jointly accounting for other 
salient features of real social networks such as clustering, 
community structure and so on. 
For this reason, we have also empirically validated our findings
running the DDM algorithm on realistic data sets.
Our preliminary experimental results confirm that real social networks
are indeed surprisingly simple to de-anonymize starting from
very limited side information.
  
\section{Model and matching algorithm\label{sec:model}}

\subsection{Basic assumptions}\label{subsec:basic}
We study the network  de-anonymization problem in the case of two social 
networks $\Gc_1(\Vc_1,\Ec_1)$ and $\Gc_2(\Vc_2,\Ec_2)$, although our model and analysis can be extended 
to the case in which more than two networks are available.
Both $\Gc_1$ and $\Gc_2$ can be fairly considered to be 
sub-graphs of a larger, inaccessible graph $\Tc(\Vc,\Ec)$ 
representing the groundtruth, i.e., the underlying
social relationships between people.
We will assume for simplicity that all graphs above have the same set of 
vertices $\Vc$ with cardinality $|\Vc|=n$, i.e., $\Vc_1=\Vc_2=\Vc$,
although 
this assumption can be easily removed by seeking to match only the intersection of vertices
belonging to $\Gc_1$ and $\Gc_2$.
We remark that $\Gc_1$ and $\Gc_2$ do not necessarily 
represent subsets of social relationships as observed in totally 
different systems (e.g., Facebook and Twitter). They could
also be obtained within the same communication system (i.e., from traces
of emails, or from traces of phone calls), due to the fact that 
users employs two ID's in the same system (i.e.,
two email addresses, or two SIM cards).  

We need a mathematical model describing how edges $\Ec_1$ and $\Ec_2$ 
are selected from the groundtruth set of edges $\Ec$.
Any such model will necessarily be an imperfect representation of 
reality, since a large variety of different situations
can occur. A user might employ either of her ID's
to exchange messages with a friend, or use only one of them
to communicate with a given subset of friends.
General, realistic models trying to capture possibly
heterogeneous correlations (positive or negative) in the
set of neighbors of a vertex as seen in $\Gc_1$ and $\Gc_2$   
become inevitably mathematically intractable. We therefore
resort to the same assumption adopted in previous  
mathematical work \cite{pedarsani,lattanzi,Grossglauser}: 
%which consists in assuming that 
each edge in $\Ec$ is retained in $\Gc_1$ (or $\Gc_2$) with a fixed probability $s$, 
independently between $\Gc_1$ and $\Gc_2$, and independently of all 
other edges\footnote{Two different probabilities 
for $\Gc_1$ and $\Gc_2$ (also different from vertex to
vertex) could be considered, provided that they do not depend on $n$.}.
%As long as all edge sampling probabilities are independent of each other, and 
%comprised in a fixed interval $[s_{\min}, s_{\max}]$ which does not depend on $n$, 
%our asymptotic results do not change.}.
This  model serves as a reasonable, first-step
approximation of real systems, which permits obtaining
fundamental analytical insights. Moreover, authors in \cite{pedarsani}
have experimentally found, by looking at temporal snapshots of an email 
network, that the above assumption %of independent edge sampling 
is largely acceptable in their case. 

Another key element is the model for the underlying social graph 
$\Tc$. To understand the impact of the power-law distribution of vertex degree, 
which surely characterizes realistic social networks, we have chosen 
a simple model known in the literature as Chung-Lu random graph \cite{ChungLu}.
In contrast to the classic model of Erd\"{o}s--R\'{e}nyi,
Chung-Lu graphs permit considering a fairly general vertex degree distribution
while preserving the nice property of independence among edge
probabilities, 
%(i.e., the existence of an edge between any two vertices 
%is independent of all other edges), 
which is of paramount importance in the analysis.

\begin{definition}
A Chung-Lu graph is a random graph of $n$ vertices 
where each vertex $i$ is associated with a positive 
weight $w_i$. Let $\bar{w} = \frac{1}{n}\sum_\Nc w_i$ be the average  weight.
Given two vertices $i,j\in \Vc$, with $i\neq j$,  
the undirected  edge $(i,j)$ is included in the graph with probability 
%\begin{equation}
$p_{ij}=\min \left \{\frac{w_i w_j}{n\bar{w}},1\right\}$,
%\end{equation}
independently of the inclusion of any other edge in $\Ec$. 
\end{definition}
To avoid pathological behavior, it is customary in the Chung-Lu model to assume 
that the maximum vertex weight is $O(n^{1/2})$.
Doing so, weight $w_i$ essentially coincides with  
the average degree of vertex $i$, i.e., $p_{ij}=w_i w_j/(n\bar{w})$. 
In our work, we will assume for simplicity
that weights are deterministic \footnote{Our results 
generalize to the case of  weights being r.v. as well.} (but note that they depend on $n$, albeit we avoid
explicitly indicating this).
%though they could be themselves 
%random variables
A suitable way to obtain a power-law degree sequence with exponent $\beta$ (with $2 < \beta < 3$, as
typically observed in real systems) is to set $w_i = \bar{w}
\frac{\beta-2}{\beta-1} (\frac{n}{i+i_0})^{1/(\beta-1)}$ where $i_0$
can be chosen such that the maximum degree is $O(n^{1/2})$.  In the
following, we will assume $\bar{w}$ to be a finite constant, although our analysis can be easily extended to the more general case in which  $\bar{w}$ scales with $n$.

\subsection{Problem definition}
The network  de-anonymization problem under study can be formulated
as follows. We assume  the underlying social network graph
$\Tc(\Vc,\Ec)$ to be a known instance of a Chung-Lu graph having 
power-law degree distribution with exponent $\beta$ (with $2 < \beta < 3$).
However, we cannot access its edge set $\Ec$.
Instead, we know the complete structure of two sub-graphs 
$\Gc_1$ and $\Gc_2$ obtained by independently sampling
each edge of $\Ec$ with probability $s$. Also, each edge in 
$\Ec$ is assumed to be (independently) sampled twice, the first time
to determine its presence in $\Ec_1$, the second time to determine its presence
in $\Ec_2$. Note that both vertices $\Vc_1$ and $\Vc_2$ must be 
considered to be assigned after a random permutation of 
indexes $1,2,\ldots,n$.
The objective is to find the correct match among them,
i.e., to identify all pairs of vertices $[i_1,i_2] \in \Vc_1\times  \Vc_2$ such that  
$i_1$ and $i_2$  correspond to the same vertex $i \in \Vc$. 

We define the graph of all possible vertex pairs as the  pairs graph
$\Mc(\Vcb, \Ecb)$, with $\Vcb =\Vc_1\times  \Vc_2$ and  $\Ecb =\Ec_1 \times \Ec_2$, 
 %as the graph in which any  vertex  corresponds to a pair of vertices
 %$[i_1,j_2]$  with $i_1\in \Vc_1 $ and  $j_2\in \Vc_2 $,  and any edge
 %in $\Ec_1 \times \Ec_2$ 
 %corresponds to a pair of edges $[e_1, e_2]$ with $e_1 \in \Ec_1$ and
 %$e_2 \in \Ec_2$.
In  $\Mc(\Vcb, \Ecb)$ there exists  an edge connecting $[i_1,j_2]$ with
 $[k_1,l_2]$  
  iff  edge $(i_1,k_1)\in \Ec_1$ 
and edge $(j_2,l_2)\in \Ec_2$.
We will slightly abuse the notation and denote
the pairs graph referring to $\Gc_1(\Vc_1,\Ec_1)$ and
$\Gc_2(\Vc_2,\Ec_2)$ by $\Mc(\Tc)$. 
Also, given two pairs $[i_1,j_2]$ and  $[k_1,l_2]$ in $\Mc(\Tc)$,
they are said to be conflicting pairs if either $i_1=k_1$ and $j_2\neq
l_2$, or $j_2=l_2$ and $i_1\neq k_1$. 
We will refer to pairs  $[i_1,i_2]$, whose vertices correspond to the same vertex $i\in \Tc$ as good pairs, and to all others (e.g., $[i_1,j_2]$) as bad pairs. The generic pair will be denoted by  $[*_1,*_2]$.

To help identifying good  pairs, 
we assume there exists a subset of a-priori matched vertices,
named {\em seed} set and denoted by $\Ac_0(n)$,
of cardinality $a_0$.
We will  consider two variants of the problem
which differs in the way seeds are assumed to be 
selected among the $n$ vertices. In the former variant, they are
assumed to be selected
at wish, but using just information on the vertex degree. 
In the latter, we assume that they can be 
 selected uniformly at random among all vertices.

\begin{algorithm}
\begin{algorithmic}[1]
%\Require $\Mc(\Tc)$
\State $\Ac_0=\Bc_0=\Ac_0(n)$, $\Zc_0=\emptyset$
%$I_r^k(u)\gets 0, 
\While{$\Ac_t\setminus \Zc_t\neq\emptyset$}  
\State $t=t+1$
\State Randomly select a pair $[*_1,*_2]\in \Ac_{t-1}\setminus \Zc_{t-1}$  
and add one  mark to all neighbor pairs of  $[*_1,*_2]$ in
$\Mc(\Tc)$.  
\State Let  $\Delta \Bc_t$ be the set of  all  neighboring pairs  of   $[*_1,*_2]$ in $\Mc(\Tc)$ whose mark counter has reached threshold 
 $r$   at time $t$.
\State Construct   set $\Delta   \Ac_t \subseteq \Delta \Bc_t $ as follows. 
Order the pairs in $\Delta \Bc_t $ in an arbitrary way,
  select them sequentially and test them for inclusion in $\Delta A_t$
\If{the selected  pair in  $\Delta \Bc_t$  has no conflicting pair in $\Ac_{t-1}$ or $\Delta \Ac_t$}
\State Insert the pair in $\Delta \Ac_t$
\Else \State Discard it
\EndIf
\State $\Zc_t=\Zc_{t-1}\cup[*_1,*_2]$, $\Bc_t=\Bc_ {t-1}\cup\Delta
\Bc_t $,  $\Ac_t=\Ac_{t-1}\cup\Delta \Ac_t$
\EndWhile 
\State \Return $T=t$, $\Zc_T=\Ac_T$
\end{algorithmic}
%\vspace{.5cm}
\caption{\label{alg:PGM}The PGM algorithm}
\end{algorithm}
\vspace{-3mm}

\subsection{Overview of the DDM algorithm} \label{sec:overview}
Before providing a high-level description of our matching algorithm (DDM),
we briefly recall the simple procedure adopted in \cite{Grossglauser}  
in the case of Erd\"{o}s--R\'{e}nyi graphs. In essence, their
algorithm, referred to as PGM (percolation graph matching), maintains a mark counter, initialized to zero, for any pair $[*_1,*_2]\in \Mc(\Tc)$
that can still potentially be matched.
The counter is increased by one 
whenever the candidate pair becomes {\em neighbor} of an already
matched pair. Two pairs $[*_1,*_2]$ and $[*_1',*_2']$ are said to be neighbors 
if 
they are adjacent on $\Mc(\Tc)$, i.e., 
 edge $(*_1,*_1')\in \Ec_1$ and edge $(*_2,*_2') \in \Ec_2$.
Among the candidate pairs whose counter is larger than or equal  
to a fixed threshold $r$, the algorithm selects one uniformly at random, 
adding it to the set of matched pairs. After this, counters 
are updated. Note that some candidate pairs might have to be permanently
discarded because they are conflicting with previously matched pairs. The algorithm 
proceeds until no more pairs can be matched. Of course seeds will 
be matched irrespective of their mark counter.
The PGM algorithm, although potentially suboptimal, is simple enough that  
its performance can be predicted using known results from bootstrap 
percolation \cite{Janson}, establishing a lower bound
on the number of seeds required to correctly match almost all 
vertices. 
A more formal description of the PGM algorithm is given in Alg.\,\ref{alg:PGM},
where: 
%\begin{itemize}
%\item 

\noindent
$\Bc_t(\Tc)$ is the set of  pairs in $\Mc(\Tc)$
  that at time step $t$ have already collected a least
$r$ marks. 
 It is composed of  $\GoodB_{t}(\Tc)$ and $\BadB_{t}(\Tc)$,  comprising good and bad pairs,
respectively\footnote{The dependency of the above sets on the generic  pairs  graph
$\Mc(\Gc)$, indicated by $\Gc$, is dropped whenever not strictly necessary.}. 
%\item 

\noindent
$\Ac_t(\Tc)$ is the set of matchable pairs at time
   $t$. 
In general, $\Ac_t(\Tc)$ and $\Bc_t(\Tc)$ do not coincide as
  $\Bc_t(\Tc)$ may include conflicting pairs that are not present in
  $\Ac_t(\Tc)$. $\Ac_t(\Tc)$ includes two subsets of
  good and bad pairs, denoted 
   by $\Good_t(\Tc)$ and $\Bad_t(\Tc)$, respectively. 
%\item 

\noindent
$\Zc_t(\Tc)$ is the set of pairs in $\Ac_{t-1}(\Tc)$ that
  have been matched at time $t$. By construction, $|\Zc_t|=t$ $\forall t$.
%\end{itemize}
%Note that, by construction, $|\Zc_t|=t$ $\forall t$; thus, $|\Ac_T|=|\Zc_T|=T$.  

In our work, since we  want to establish lower bounds
on the number of seeds by means of bootstrap percolation theory, we keep
the simplicity of the PGM algorithm, adding some fundamental
improvements to exploit the heterogeneity of vertex degrees.
Before explaining our approach, we make the following 
observations on the PGM algorithm described above for
Erd\"{o}s--R\'{e}nyi graphs. First, pairs are selected
irrespective of the degree of their constituting vertices.
Intuitively, in Erd\"{o}s--R\'{e}nyi graphs this is not so important,
since vertices degree (which is binomial distributed) 
is highly concentrated around the mean, and all matchable pairs are 
essentially equivalent. Second, there exists a unique threshold $r$, 
common to all pairs, which is a fixed parameter of the algorithm 
subject to the constraint $r \geq 4$.

Our DDM algorithm for power-law graphs is based instead
on partitioning the vertices on the basis of their degree. It then implies 
a careful expansion of the set of matched pairs through
the various partitions, using also different thresholds
and seed sets at the various stages of the process. 

In particular, we first isolate a specific slice (i.e.,  sub-graph of
$\Mc(\Tc)$), $\Mc_1$,  induced by  vertices
having large (but not too large) degree. 
$\Mc_1$ includes pairs whose vertices have weights between 
$\alpha_1=n^{\gamma}$ and $\alpha_{2}=n^{\gamma}/2$,
where $\gamma$ is a constant (slightly) smaller than $1/2$.
This slice is somehow the crucial one: we show that its percolation   
 triggers the entire matching process, as the identification 
of all other vertices in the network follows easily after we correctly 
match all pairs in $\Mc_1$.
Note that degrees of vertices in $\Mc_1$ are fairly homogeneous (a constant factor
of difference), so that the results for
Erd\"{o}s--R\'{e}nyi graphs can be adapted to  this slice.

Vertices having degree smaller than those in $\Mc_1$
are partitioned in geometric slices $\Mc_k$ including vertex pairs with  
weights between $\alpha_{k}$ and $\alpha_{k+1} = \alpha_{k}/2$,
with $k\ge 2$. Then, a top-down cascading process is unrolled starting from $\Mc_1$, 
where matched pairs in a slice are used as seeds to identify the
good pairs in the slice below, and so on.

Vertices with very large degree are identified at the end, 
using as seed set a properly defined subset of 
previously matched pairs with relatively small degree.

Here we have provided just the basic idea of our DDM algorithm: 
many subtleties must be addressed to show its correctness.  
Among them, we emphasize the problem that the DDM algorithm has no
direct access to vertex weights (i.e., it does not know the original 
degree of a vertex in $\Tc$), and can only make use of the observable 
vertex degrees in $\Gc_1$ and $\Gc_2$. As a consequence, in the Appendix
we  show that our matching algorithm is sufficiently robust
also in the presence of imperfect (estimated) vertex partitioning. 
  
At last, we remark  that,  when $\bar{w}$ is finite,   a finite fraction of good pairs may not be identified, no matter which matching algorithm is used. This fact  can be immediately grasped by observing that any  good pair $[i_1, i_2]$  can  be identified  only if  both $i_1$ and $i_2$ have at least $r$ neighbors in $\Gc_1$ and $\Gc_2$. Clearly, due to independent edge sampling, a finite fraction of vertices in $\Tc$ that have bounded degree  gives  origin to vertices with degree smaller than $r$ in either $\Gc_1$ or $\Gc_2$.

%\section{Related work} \label{sec:related}

\section{Notation and preliminary results\label{sec:our-pre-res}}
We first recall   the results on Erd\"os-R\'enyi graphs obtained in \cite{Grossglauser}.
In particular, one of the major results that we will use in  our analysis is given by the
theorem below~\cite[Th.\,1]{Grossglauser}.
\begin{teorema} \label{theorem:gross}
Let the groundtruth graph  be an  Erd\"os-R\'enyi  random graph $G(n,p)$.
Let  $r\ge 4$.
Denoted with: 
\begin{equation} \label{ac}
% t_c=\left( \frac{(r-1)!}{nq^r}\right)^{\frac{1}{r-1}}  \quad
 a_c=\left(1-\frac{1}{r}\right) \left( \frac{(r-1)!}{n(ps^2)^r}\right)^{\frac{1}{r-1}}\, .
\end{equation}

For $n^{-1} \ll ps^2 \le s^2 n^{-\frac{4}{r}}$, we have:
%i) if $lim_{n\to\infty } a_o/a_c <1$, the PGM algorithm stops with $a_*<\frac{r}{r-1}a_0$ with high probability (w.h.p.). 
that, if  $a_o/a_c\to a>1$, the PGM algorithm matches a number of good pairs 
equal to $|\Good_T|=n-o(n)$ w.h.p.
Furthermore, $\Bad_{T}=\emptyset$ w.h.p. 
\end{teorema}
Observe that under the  assumptions of Theorem \ref{theorem:gross}, we
have  $T=|\Ac_T|=|\Good_T|=n-o(n)$. 
The  two corollaries below, which  can be derived from the arguments presented in~\cite{Grossglauser},  
 strengthen the result in Theorem~\ref{theorem:gross}  
%in the supercritical
% regime under the additional condition $p \gg \sqrt{\frac{n^{-3/r -1}}{ s^2}}$.
%provide useful complementary results that 
and  will  come in handy in the following.
\begin{corollario} \label{corollary:gross1}
For    any    $\epsilon>0$,  define
$t_0=\min\left ( T,\frac{n^{-3/r-\epsilon}}{(p s)^2}\right )$.  Then,  
$\BadB_{t_0}=\emptyset $ w.h.p. 
\end{corollario}
When $t_0=T$, the corollary guarantees that $\Bad_T \subseteq \BadB_T =\emptyset$,
%=\BadB_{t_0}
i.e., no bad pairs are matched by the PGM algorithm. 
When $t_0<T$ (i.e., for  $p \gg \sqrt{\frac{n^{-3/r -\epsilon-1}}{ s^2}}$),  we
complement the above statement with the corollary below.
%Note that, for simplicity, we upper bound 
%the expression $\sqrt{n^{-3/r -\epsilon-1}}$ with $\sqrt{n^{-3/r-1}}$
%(i.e., by setting $\epsilon=0$), and  use the latter in the following.
\begin{corollario} \label{corollary:gross2}
%Assume $r\ge 4$, $n^{-1} \ll ps^2 \le s^2 n^{-\frac{4}{r}}$ and $a_o/a_c\to a>1$.  
Under the  conditions of  Theorem~\ref{theorem:gross}, 
for  $p \gg \sqrt{\frac{n^{-3/r -1}}{
    s^2}}$, 
 let  $t_0=\frac{n^{-3/r-\epsilon}}{(p s)^2}$  for any
 $0<\epsilon<\frac{1}{r}$. Then,   
$|\GoodB_{t_0}| = n  $    w.h.p.
\end{corollario}
%\proof Since $p \gg \sqrt{\frac{n^{-3/r -1}}{ s^2}}$  for any $ 0 < \epsilon
%< \frac{1}{r}$, we have  $t_0 \ll n$. Also,  
%since $T=n-o(n) $ by hypothesis, then $t_0<T$,  hence $\GoodB_{t_0}$ is a well defined quantity.
%Furthermore, observe that, since by Lemma \ref{lemma:gross1}
%$\BadB_{t_0}=\emptyset $ w.h.p.  (i.e.,   $\Bc_{t_0}$  comprises only good pairs),
%it descends that   $\Zc_{t_0}$  includes only good pairs  w.h.p. Indeed,  by construction,  $\Zc_{t_0} \subseteq \Ac_{t_0-1} \subseteq  \Ac_{t_0} \subseteq \Bc_{t_0}$. 
%Thus,
%\begin{multline}
% \P(|\GoodB_{t_0}| = n )=  \P \left ( \sum_i \ind_{\{
%   \text{Marks}_{[i_1,i_2]}(t_0)> r\}}=n \right ) \\
%=1-\P \left (  \cup_i ( \ind_{\{ \text{Marks}_{[i_1,i_2]}(t_0)\le r \}
%} = 1) \right)  \\
% \ge 1- \sum_i \P \left(  \ind_{\{ \text{Marks}_{[i_1,i_2]}(t_0)\le r
%   \}} = 1 \right ) =\\ 1-\sum_i \P ( \text{Bi}(n,ps^2)\le r)\\
%\ge 1- n \exp(r- t_0 ps^2) \to 1
%\end{multline}
%where the last inequality comes from the application of the Chernoff bound to $\P( \text{Bi}(n,ps^2)\le r)$, while $n \exp(r- t_0 ps^2) \to 0$ because by   hypothesis   $p \le n^{-\frac{4}{r}} \ll n^{-3/r-\epsilon}/\log n$  .
%\endproof

%Note that Lemmas \ref{lemma:gross1} and \ref{lemma:gross2} together allow to slightly strengthen the statement of Theorem~\ref{lemma:gross}  in the supercritical
% regime under the additional condition $p \gg \sqrt{\frac{n^{-3/r -1}}{ s^2}}$. Indeed,  the fact that 
The fact that,  
 for some $t_0<T$,   $|\GoodB_{t_0}| = n  $ and $\BadB_{t_0}=\emptyset $  jointly occur 
  w.h.p. implies that  the 
PGM   algorithm matches almost all the good pairs (i.e., $|\Good_T|=n$
and $\Bad_T=\emptyset$) w.h.p. 
This is because, by construction,  
$\Good_{t_0} =  \GoodB_{t_0}$. Indeed,  $\GoodB_{t_0}$ contains no conflicting pairs and none of the pairs in $ \GoodB_{t_0}$ can be blocked by 
previously  matched bad pairs  since   $\BadB_{t_0}=\emptyset$.
%Lastly, note that, since $|\Ac_T|=|\Good_T|=n$ and $|\Ac_T|=T$, it
%follows that $T=n$. 

We now extend the above results to  Chung-Lu graphs.
First we   introduce the key   concept of {\em increasing} property. 
%In the rest of the paper, we consider the ground truth graph $\Tc(\Vc,\Ec)$ to
%be a Chung-Lu random graph.  
%It is straightforward to see that, since $\Tc(\Vc,\Ec)$  is a  Chung-Lu random graph, 
% $\Gc_1$ and $\Gc_2$ are Chung-Lu graphs as well. Also, recall that
%$\Mc(\Tc)$ is the matching graph including all possible vertex
%pairs $[i_1,j_2]$, with $i_1 \in \Gc_1$ and $j_2 \in \Gc_2$. 
%We make the following key observations. 

Let $\Hc(\Vc,\Ec_H)$ and $\Kc(\Vc,\Ec_K)$  be two random graphs insisting on the same set of vertices $\Vc$, where $\Ec_H \subseteq  \Ec_K$, i.e.,  
 $\Ec_H$  can be obtained by sampling $\Ec_K$.  We  define the following  partial order relationship:
$\Hc(\Vc,\Ec_H) \le_{st} \Kc(\Vc,\Ec_K)$. Then,  we can define a  vertex property   $\Rc$  satisfied by a subset of  vertices, and  denote with 
 $ \Rc (\Hc) \subseteq \Vc$  the set of  vertices  of
 $\Hc$  that satisfy   property $\Rc$. 
We say that $\Rc$ is {\em monotonically increasing with respect to the graph ordering relation ``$\le_{st}$''} 
if $\Rc(\Hc) \subseteq \Rc(\Kc)$  whenever $\Hc\le_{st} \Kc$.

In our case, for any $0\le t\le T$,  sets  $\Bc_{t}$, $\GoodB_{t}$,  $\BadB_{t}$
are all  monotonic increasing with  respect to  relationship
``$\le_{\text{st}}$'' defined on the pairs graph $\Mc(\Tc)$.    
%Instead, nothing can be said on $\Ac_t$, $\Good_t $ and
%$\Bad_t$  due to the effect of mutual conflicts among pairs (i.e., the presence of a
%pair in $\Ac_t$ prevents the further addition of all conflicting pairs
%in $\Bc_t$).
Below, we use this observation and show that a properly defined subgraph of a Chung-Lu  graph
 can be lower and upper bounded (w.r.t. ``$\le_{\text{st}}$`` relation) by  Erd\"os-R\'enyi
graphs. Secondly, considering the above subgraph $\Gc_0$, we prove that the
pairs graph  $\Mc(\Gc_0)$ can be lower and upper bounded by 
properly defined Erd\"os-R\'enyi
graphs. %(Proofs are omitted for brevity; they can be found in \cite{techrep}.)
\begin{proposizione} \label{carabinieriprop}
Given a   Chung-Lu  random graph $\Tc(\Vc,\Ec)$,  for any given
interval of vertex weights $[w_{\min}, w_{\max} ]$, we define: 
 $\Vc_0 \subseteq \Vc$,  $\Vc_0 =\{i \in \Vc | w_i \in [w_{\min}, w_{\max} ]\}$  with  $|\Vc_0|=n_0$ and
 $\Ec_0=\{(i,j) \in \Ec | i,j\in \Vc_0 \}$.
Now, consider   $\Gc_0=(\Vc_0,\Ec_0)$, i.e.,  the sub-graph of
$\Gc_T(\Vc,\Ec)$ induced by only  vertices in $\Vc_0$. The following
relationship holds:  
$ G(n_0,p_{\min})  \le_{st}  \Gc_0 \le_{st} G(n_0,p_{\max})$, with
$G(n_0,p_{\min})$ and $G(n_0,p_{\max})$ 
 being Erd\"os-R\'enyi graphs and  $p_{\min}=w^2_{\min}/(n\bar{w})$ and $p_{\max}=w^2_{\max}/(n\bar{w})$.
\end{proposizione}
\proof The proof is immediate in light of the fact that every edge $(i,j)$, with $i,j\in \Vc_0$, by construction belongs to  $\Ec_0$ independently of other edges with a probability 
$p_{\min} \le p \le p_{\max}$.
\endproof
\begin{proposizione} \label{LB-UB-matching}
Given the above   Chung-Lu  subgraph $\Gc_0(\Vc_0,\Ec_0)$ and
the Erd\"os-R\'enyi graphs $G(n_0,p_{\min})$ and $G(n_0,p_{\max})$,
consider the two graphs obtained from each of them by
independent edge sampling with  probability $s$.
$\Mc(\Gc_0)$, $\Mc(G(n_0,p_{\min}))$ and $\Mc(G(n_0,p_{\max}))$
are the corresponding pairs graphs. If $ G(n_0,p_{\min})  \le_{st}  \Gc_0
\le_{st} G(n_0,p_{\max})$, then 
$ \Mc(G(n_0,p_{\min})) \le_{st} \Mc(\Gc_0) \le_{st} \Mc(G(n_0,p_{\max}))$. 
\end{proposizione}
\proof By construction, if the ``$\leq$'' relationship holds for $\Gc_0(\Vc_0,\Ec_0)$, $G(n_0,p_{\min})$ and $G(n_0,p_{\max})$,
then it holds also for the graphs obtained from them by
independent sampling. Thus, by construction, it is also valid for the
corresponding matching graphs. 
\endproof

Next,   we  present our first main result,  which shows that the PGM
algorithm can  successfully match all good pairs in a subgraph 
$\Gc_0$, of a Chung-Lu graph.  
%\textcolor{red}{The core of the proof consists in applying Propositions
%\ref{carabinieriprop} and \ref{LB-UB-matching}. Then, applying the previous results on the
%Erd\"os-R\'enyi graphs that lower and upper bound $\Gc_0$, 
%by monotonicity we prove our thesis.} 
\begin{teorema} \label{propLU}
Consider   $\Gc_0$ obtained from  $\Tc$ as
defined in Proposition \ref{carabinieriprop}.
Applying the PGM algorithm on $\Mc(\Gc_0)$  guarantees that $|A_T(\Gc_0)|=n_0 $ and
$\Bad_{*}(\Gc_0)=\emptyset$ w.h.p., provided that:
\begin{enumerate}
\item $n_0\to \infty$ as $n\to \infty$; 
\item  $p_{\min}=w^2_{\min}/(n\bar{w})$  satisfies:    $p_{\min}  \gg \sqrt{\frac{n^{-3/r-1}}{ s^2}}$;
\item  $p_{\max}=w^2_{\max}/(n\bar{w})$  satisfies: $ p_{\max} \le  n_0^{-\frac{4}{r}}$;
\item $\lim_{n\to \infty} a_o/a_c>1$  with $a_c$ computed  from (\ref{ac}) by   setting $p=p_{\min}$.
%\item $a_o/a^2_c>1$  with $a^2_c$ computed as in Lemma \ref{lemma:gross} for $p_{\max}$.
\end{enumerate}
%%% PEZZO SOTTO DA REINTRODURRE SE SLICE IMPERFETTE -- begin 
Under conditions 1)-4),  the PGM  successfully matches w.h.p. all the  correct  pairs  (with no errors)
also in any  subgraph $\Gc_0^{'}$ of $\Gc_0$  that comprises a finite
fraction of vertices of  $\Gc_0$  and all the edges between the selected vertices.
%%% -- end
\end{teorema}
\proof  First observe that, if we find $t_0$ with $t_0=o(n_0)$ such that $\BadB_{t_0}(\Gc_0)=\emptyset$ w.h.p., then  we have w.h.p that $\forall t\le t_0$:
\begin{align} \label{neweq}
|\Ac_t (\Gc_0)|= & | \GoodB_t (\Gc_0) |\stackrel{(a)}{\ge}  \nonumber \\
 | \GoodB_t (G(n_0, p_{\min}) )| \stackrel{(b)}{=} & | \Ac_t (G(n_0, p_{\min}))| \stackrel{(c)}{>}t. 
\end{align}
 In (\ref{neweq}),  inequality (a) holds by monotonicity of sets  $\GoodB_t $  with respect  to ``$\le_{\text{st}}$'',  
while   equality (b)  descends from Theorem \ref{theorem:gross}.  Inequality (c)  descends from the following argument. Denoted by 
$T_G=\min\{ t \,\mbox{\em s.t.} \, |\Ac_t (G(n_0, p_{\min})|=t\}$, by Theorem~\ref{theorem:gross} we have 
$T_G=n_0-o(n_0)$. Since $t_0=o(n_0)$, $t_0<T_G$, i.e., $| \Ac_t (G(n_0, p_{\min}))| > t$   for $t\leq t_0$. 
From (\ref{neweq}),  
we immediately get $t_0< T$, with  $T=\min\{ t \,\mbox{\em s.t.} \, |\Ac_t (\Gc_0)|=t\}$.

Now, let us define,  for an arbitrarily small   $\epsilon>0$, $t_0=\frac{n_0^{-3/r-\epsilon}}{(p_{\max} s)^2}$; observe that  by construction  $t_0=o(n_0)$.  
We prove that  $\BadB_{t_0}(\Gc_0)=\emptyset$ exploiting the monotonicity of  $\BadB_{t_0}$ with respect to 
``$\le_{\text{st}}$''.  
Indeed, $|\BadB_{t_0}(\Gc_0)|\leq |\BadB_{t_0}(\Gc(n_0, p_{\max}))|$, with  $\BadB_{t_0}(\Gc(n_0, p_{\max}))= \emptyset $ w.h.p.
 as  immediate consequence of  Corollary \ref{corollary:gross1}
(recall that   $n_0\to \infty$ as $n \to \infty$).
Furthermore,
by Corollary \ref{corollary:gross2}, for an  arbitrary
$0<\epsilon'<1/r$,  define  
$t_1=  \frac{n_0^{-3/r-\epsilon'}}{(p_{\min} s)^2}=o(n_0)$. We have:  $|\GoodB_{t_1}(\Gc(n_0,p_{\min})|=n_0$. 
Next,  by monotonicity, we have  $|\GoodB_{t_1}(\Gc_0)|\ge
|\GoodB_{t_1}(\Gc(n_0, p_{\min})|=n_0$, provided that $t_1\le T$.

At last, since $p_{\max}/p_{\min}=K^2$,  we can always choose an
 $\epsilon<\epsilon'$  such that  
$T>t_0=\frac{n_0^{-3/r-\epsilon}}{(p_{\max} s)^2} >\frac{n_0^{-3/r-\epsilon'}}{(p_{\min} s)^2}=t_1$. 
 Thus,  since $\GoodB_{t_0}(\Gc_0)$ is by construction non decreasing
 with $t$, we have:   $|\GoodB_{t_0}(\Gc_0)|\ge
 |\GoodB_{t_1}(\Gc_0)|=n_0$. In conclusion,  there exists a $t_0<T$
such that  $|\GoodB_{t_0}(\Gc_0)|=n_0$  and  $\BadB_{t_0}(\Gc_0)=\emptyset$.  Hence,  $|\Good_{T}(\Gc_0)|=|\Good_{t_0}(\Gc_0)|=|\GoodB_{t_0}(\Gc_0)|=n_0$ 
and $|\Bad_{T}(\Gc_0)|=|\BadB_{t_0}(\Gc_0)|=0$. 
%%% PEZZO SOTTO DA REINTRODURRE SE SLICE IMPERFETTE -- begin 
The extension of previous results to $\Gc_0^{'}$ is immediate in light of the fact that $\Gc_0^{'}$ inherit all properties of  $\Gc_0$ 
%%%%%%%%% --- end
\endproof

%%% PEZZO SOTTO DA REINTRODURRE SE SLICE IMPERFETTE -- begin 
%\begin{comment}  
The following corollary immediately follows:
\begin{corollario} \label{propLU-1}
Under same conditions of Theorem \ref{propLU}, the DDM algorithm  can be 
successfully applied  to an imperfect slice $\Mc'(\Gc_0)$  (with $\Mc'(\Gc_0) \subset \Mc(\Gc_0)$)   
comprising a finite fraction of the pairs in $\Mc(\Gc_0)$ and satisfying  the following constraint: 
a bad $[i_1,j_2] \in \Mc(\Gc_0) $  is  included in $\Mc'(\Gc_0)$
only if either    $[i_1,i_2]$ or $[j_1,j_2]$ are in also $\Mc'(\Gc_0)$.
\end{corollario}
\proof
Essentially the scheme of previous Theorem can be repeated to show that there exists  $t_1<T$ 
  such that $\Bc_{t_1}(\Mc'(\Gc_0))$ comprises all the good pairs in
  $\Mc'(\Gc_0)$  and no bad pairs.

Indeed, first observe that  $\Mc'(\Gc_0)$ can be always  transformed into  $\Mc(G'_0)$ for some $G'_0$ by adding and removing only bad pairs;
second, from Theorem~\ref{propLU} we know that,   
for $t_1=\frac{(n'_{0})^{-3/r-\epsilon}}{(p_{\min} s)^2}=o(n'_{0})$, 
it holds:
 $\GoodB_{t_1}(\Mc (\Gc'_0))=n'_{0}$  where $n'_{0}$ denotes the number of vertices in $G'_0$ 
(equal, by construction, to the number of good pairs in  $\Mc'(\Gc_0)$). 
Third, again from Theorem~\ref{propLU}, it holds that $\BadB_{t_1}(\Mc (\Gc'_0))=\emptyset$.
 
Hence, if we prove that $\BadB_{t_1}(\Mc'(\Gc_0))=\emptyset$, we can conclude that $\GoodB_{t_1}(\Mc'(\Gc_0))=n'_{0}$ since  under the condition $\BadB_{t_1}(\Mc'(\Gc_0))=\emptyset$
 necessarily sets $\GoodB_t(\Mc'(\Gc_0))=\GoodB_t(\Mc(\Gc'_{0})) $  for every  $t\le t_1$  (this because by construction the   subgraphs of 
of $\Mc'(\Gc_0)$ and $\Mc(\Gc'_{0})$ induced by their good pairs  are identical by construction).
To  prove that   $\BadB_{t_1}(\Mc'(\Gc_0))=\emptyset$, we can repeat
the same arguments as in the proof of Corollary \ref{corollary:gross1}. Indeed  we can upper-bound the number 
of marks collected at time $t$ by every bad pair $[i_1,j_2]\in \Mc'(\Gc_0)$  with a r.v. Bi$(t, p^2_{\max} s^2)$ and, then, operate
exactly  as in the proof of Lemma \ref{corollary:gross1} to show that   $\PP\{\BadB_{t_1}(\Mc'(\Gc_0))\neq \emptyset\}\to 0$.

 The assert then follows from the observation that, given the
 constraint stated in the corollary,    
this is enough conclude that no bad pairs can be  matched  for
$t>t_1$, because they will necessarily being blocked by a previously matched good pair.

\endproof
%\end{comment} 
%%% PEZZO SOTTO DA REINTRODURRE SE SLICE IMPERFETTE -- end

\section{DDM algorithm and analysis\label{sec:problem}}

Here we present the details of the DDM algorithm and 
prove the following main results.

\noindent 
(i) For a sufficiently large seed set, the DDM algorithm successfully matches $\Theta(n)$ good pairs and no bad pairs. Also, it matches all good pairs (except for a negligible fraction)  constituted by 
vertices with sufficiently high weight, i.e., that tends to infinity as $n\to \infty$.

\noindent
(ii) The above results holds for a seed set as small as $n^\epsilon$ (with any arbitrary $\epsilon >0$) when the seeds can be chosen based on the vertices degree. When instead the seeds are uniformly distributed over the graph, 
$n^{\frac{1}{2}+\epsilon}$ seeds are necessary. 

\noindent
(iii) In the more general case where seeds are arbitrarily distributed, the key parameter for triggering the good pairs identification process 
is represented by  the size of the set of edges between the seeds and the rest of pairs in the graph.

%\subsection{Slicing the graph\label{subsec:slice}}

%Without loss of generality, we assume that the weights of the vertices
%in our  ground-truth,  Chung-Lu, random graph,  $\Tc(\Vc,\Ec)$,  are  deterministically assigned.
%Furthermore, we set the  weight of the generic vertices
%  $i_1 \in \Vc_1$ and $i_2 \in \Vc_2$  equal to $s^2\w_i$, with
% $w_i$ being the weight of the corresponding vertex $i \in \Vc$. 
%Recall that in Chung-Lu graphs the maximum weight is $O(n^{1/2})$~\cite{ChungLu} and that 
% weight $w_i$ corresponds to the average degree of vertex $i$.
%In particular, we assume that  $w_i=\min \left [ (n^{-1} i )^{\frac{1}{1-\beta}},
%n^{1/2} \right ]$, with $2<\beta<3$. This implies that the empirical
%distribution of 
%the vertex weights follows a power law with parameter $\beta$. Also, we
%denote by $\bar{w}$ the average value of the vertex weights, which is
%finite by construction. 

We start by generalizing  the approach proposed in \cite{Amini}. 
We slice the pairs graph $\Mc(\Tc)$ 
 into subgraphs $\Mc_x$, of  pairs of vertices with weight
comprised between thresholds $\alpha_{x}$ and $\alpha_{x+1}$ ($x\in {\mathbb{N}}$).
By doing so, we assume the vertices weights  to be directly
accessible by the DDM algorithm. In practice, this 
is not possible: the DDM algorithm has direct access only to vertex degrees on $\Gc_1$ and  $\Gc_2$. 
%In other words,  slices of $\Mc(\Tc)$ can be  obtained only resorting
%to the information about the actual degree of  vertices in $\Gc_1$
%and $\Gc_2$.  
 In the Appendix, %\ref{app:slices}, 
we present a technique to work around  this issue and relax the above assumption.
%Furthermore, we remark that at every step of the DDM algorithm,  the 
%  threshold $r$ and the seed set are
%properly determined for each slice we consider. 
%The fact that such parameters may vary from a slice to another allows
%us to find less stringent conditions on the values they shall take.

Slices of the pairs graph are constructed as follows:
%\begin{itemize}

\noindent
(i) $\Mc_0$ including pairs whose vertices have weights between $\alpha_0=n^{1/2}$ and $\alpha_{1}=n^{\gamma}$,  with  $0<\gamma<1/2$ to be determined;

\noindent
(ii) $\Mc_1$ including pairs whose vertices have  weights between $\alpha_1=n^{\gamma}$ and $\alpha_{2}=n^{\gamma}/2$;

\noindent
(iii) $\Mc_k$ including vertex pairs with  weights    between $\alpha_{k}$ and $\alpha_{k+1}$, with $k\ge 2$,  $\alpha_{k}=\alpha_{k-1}/2$,   $\alpha_{k}> 
\left (\frac{8 \bar{w}
\log n}{Cs^2(1-\epsilon)^2}\right)^{\frac{1}{3-\beta}}$ for some $\epsilon>0$;

\noindent
(iv)  $\Mc_h$ including  vertex pairs with weights   between
  $\alpha_{h}$ and $\alpha_{h+1}$, with $\alpha_{h}=\alpha_{h-1}/2$
  and 
  $\alpha_{h} \le \left (\frac{8 \bar{w}
\log n}{Cs^2(1-\epsilon)^2}\right)^{\frac{1}{3-\beta}}$ but $\alpha_{h}\rightarrow \infty$ as $n \rightarrow\infty$;

\noindent
(v) $\Mc_q$ including vertices with  weights   between
  $\alpha_{q}$ and $\alpha_{q+1}$, with  $\alpha_{q}=\alpha_{q-1}/2$ and $\limsup \alpha_{q}<\infty$. 
%\end{itemize}

Since  $\Mc_0$  is populated by only few vertices
  which are highly interconnected, starting the vertex matching procedure
  from there would likely lead to errors. We therefore   
start from $\Mc_1$  
%whose pairs include vertices
%with weight between $\alpha_1$ and $\alpha_2$, 
and then  process the vertices in $\Mc_0$ at the end by exploiting their edges with lower degree vertices.

Our goal is to show that the process of good pair matching
percolates on $\Mc_1$  faster than bad pairs, provided
that a sufficient fraction of good 
pairs  have been initially identified  (seed set).
We denote by $\Ac_0^1$  the seed set in  $\Mc_1$.
To  prove the above statement, we apply Theorem \ref{propLU},
verifying that  all its assumptions are satisfied.
\begin{proposizione} \label{prop:M1}
Good pairs are successfully matched in $\Mc_1$ 
%(as well as on $\Mc_1'$ satisfying the assumptions of Corollary \ref{propLU-1})
if the following conditions are jointly satisfied: 
$\frac{1}{4}-\frac{3}{2r} <\gamma< \frac{1}{\beta-1}$,  $r\ge \frac{4[1+\gamma(1-\beta)]}{1-2\gamma}$ 
and $ |\Ac_0^1|\gg n^{\frac{(1- 2 \gamma)r+ \gamma(\beta-1)-1}{r-1}}$.  
\end{proposizione}
\proof First, we compute the  number of good pairs  in $\Mc_1$,
denoted by $N_1$, and  make sure that $N_1$ grows to infinity when $n\to
\infty$ (as requested by condition 1) of  Theorem \ref{propLU}. We have:
\[
N_1=\sum_{i\in\Vc } \ind_{\{w_i\in [\alpha_2,\alpha_1]\}} \approx \int_{\alpha_2}^{\alpha_1} n x^{-\beta} \dd x= C n^{1+\gamma(1-\beta)} 
\]
where $C$ is a proper constant term. Clearly, $N_1\to \infty$ provided that  
$1+\gamma(1-\beta)>0$, i.e.,
$\gamma<\frac{1}{(\beta-1)}$.
Now, probabilities $p_{\min}$ and $p_{\max}$, 
 defined as  in Theorem \ref{propLU},
 satisfy the following relationship:
\[
p_{{\min},{\max}}=\Theta\left( \frac{n^{2\gamma}} {n\bar{w}} \right )= \Theta(n^{2\gamma-1}).
\]
 To  verify  condition 2)  in  Theorem \ref{propLU},   we must enforce: 
$-\frac{3}{2r}-\frac{1}{2}<{2\gamma}-1$, thus $\gamma > \frac{1}{4}-\frac{3}{4 r}$,
and to verify  condition 3)  (i.e., $p_{\max} <N_1^{-\frac{4}{r}}$),
we must have:  
$n^{2\gamma-1}\leq n^{[1+\gamma(1-\beta)]4/r}$
or, equivalently, 
\begin{equation}
r \geq \frac{4[1+\gamma(1-\beta)]}{1-2\gamma} .
\end{equation}
Next, 
%adopting the same   notation as in Theorem \ref{propLU}, 
we observe that:
\[
 a_1^c(N_1)= \left (  1- \frac{1}{r}\right) \left( \frac{(r-1)!}{N_1
     p_{\min}^r} \right )^{1/(r-1)}
\hspace{-1cm}=\Theta( n^{\frac{(1- 2 \gamma)r+ \gamma(\beta-1)-1}{r-1}}).
\]
Thus, condition 4) of Theorem \ref{propLU} is surely  satisfied if  $ |\Ac_0^1|\gg n^{\frac{(1- 2 \gamma)r+ \gamma(\beta-1)-1}{r-1}}$.
\endproof
The above is one of our main results. Essentially it states
that we can chose any $\frac{1}{4}\le \gamma < \frac{1}{2}$ and determine
 a minimal $r$ and a minimal $|\Ac_0^1|$  for which Proposition
 \ref{prop:M1} holds. 
Also,  if our goal is to minimize  $|\Ac_0^1|$, 
 $\gamma$ should be chosen as close as possible to
 $\frac{1}{2}$  (i.e., $\gamma=\frac{1}{2} -\epsilon$  for some small
 $\epsilon$). Under such condition and for a sufficiently large
 $r$, we can make the seed set arbitrarily small and still 
 correctly match all pairs.

We now consider slice $\Mc_k$ ($k>1$) and prove that: (i) the process of matching good pairs successfully propagates
from one slide to another and (ii) no errors are made.
 To this end, we first 
 look at  the number of edges from the good pairs in a slice
toward those in the slice above and 
show that the
probability that this number is smaller or equal to a threshold
goes to 0 sufficiently fast. 
We remark that in this case it is important to explicitly find  the
minimum value of $n$  for which the above result hold. Indeed, later  we have  
to  show that  similar properties  hold  uniformly over all the considered slices, for sufficiently large $n$.
\begin{teorema} \label{lemma-prop-thorugh-strips}
Consider the good pairs   $[i_1,i_2]\in \Mc_{k}$, with  vertex weight $w_i\in
[\alpha_{k+1}, \alpha_{k}]$.   Given a generic pair
$[i_1,i_2]\in\Mc_{k}$, for any $\epsilon>0$, 
  with probability  greater than $1- n^{-2}$, 
the number of its  neighboring good pairs $[l_1,l_2] \in \Mc_{k-1}$ 
 is greater than  $\rho_k= \max(4, \frac{(\alpha_k)^{4-\beta}}{\sqrt{n}}) $, 
as long as $ \left (\frac{8 \bar{w}
\log n}{Cs^2(1-\epsilon)^2}\right)^{\frac{1}{3-\beta}}=\alpha_k^*<\alpha_k<n^{\gamma}$  (with $1/4<\gamma<1/2$), and 
$n>n_1=\max\left\{ \exp \left [ \left(\frac{8
   \bar{w}
}{Cs^2}\right )^{2-\beta} \epsilon^{\beta-3}\right ],
\left (\frac{2 \bar{w}
}{Cs^2\epsilon}
\right )^{\frac{2}{1-2\gamma}} \right \}$. 
Furthermore, the  above property 
holds uniformly over the good pairs in $\Mc_{k}$ with a probability
greater than $1- n^{-1}$, under the same conditions as before on $\alpha_k$ and $n$.
\end{teorema}
\proof Given  a pair $[i_1,i_2]\in \Mc_{k}$,  for any pair
$[l_1,l_2]\in \Mc_{k-1}$,  
we denote with   $\ind_{i,l}$  the indicator function associated to the presence of an edge 
between $  [i_1,i_2]$ and $[l_1,l_2]$ in   $\Mc(\Tc)$. Note that
$\EE[\ind_{i,l}]\ge \frac{\alpha_{k+1}\alpha_{k}s^2}{n\bar{w}}=
\frac{\alpha_{k}^2s^2}{2 n \bar{w}
}= p_{min}$, and that  
$\ind_{i,l}$'s are independent r.v. Thus, by denoting  the number of
good pairs in $\Mc_{k-1}$ with 
$N_{k-1}=C n \alpha_k^{(1-\beta)}$,   %{\bf costante $C$ da def}  
and defining $\mu= N_{k-1} p_{\min}= C n s^2 \alpha_{k}^{1-\beta}\frac{\alpha_{k}^2}{2n \bar{w}
}= \Theta\left( s^2(\alpha_{k})^{3-\beta}\right) $, for any $\rho_k<\mu $, we have:
\begin{eqnarray}
 \P\left( \sum_{l\in \Mc_{k-1}}  \ind_{i,l} \leq \rho_k\right) &<& 
\P(\text{Bi}(N_{k-1}, p_{\min}) \leq \rho_k) \nonumber\\
&\le& \exp(- \delta^2 \mu/2 )
\end{eqnarray}
with $\delta= \frac{\mu-\rho_k}{\mu}$.  
In the above derivation,  the first inequality descends from the fact
that  $\sum_{[l_1,l_2]\in \Mc_{k-1}}  \ind_{i,l}$ can be stochastically
lower bounded  by a sum of $N_{k-1}$ independent Bernoulli r.v.  with average $p_{\min}$, 
while the second descends from the Chernoff bound.
Now, let us fix $\rho_k=\max\left (4,
\frac{(\alpha_k)^{4-\beta}}{\sqrt{n}}\right)= o(\mu).$ 
For any $\epsilon>0$ and  choosing $\delta =1 -\epsilon$, we have that whenever $\rho_k<(1-\delta)\mu=\epsilon \mu$,
% i.e. for   $n>\max\Big(\exp(\frac{1}{\epsilon}),\frac{2\bar{w}}{Cs^2\epsilon^2} \Big)$ :
%(i.e. when $n$ is  sufficiently large such that $\alpha_k>\frac{8\bar{w}}{Cs^2\epsilon}$  for $\alpha_k< (4\sqrt{n})^{\frac{1}{4-\beta}}$ 
% and $n^{1/2}/\alpha_k>(\frac{2\bar{w}}{Cs^2\epsilon})$ otherwise):
\[ 
  \P\left( \sum_{[l_1,l_2]\in \Mc_{k-1}}  \ind_{i,l} \leq \rho_k\right) <\exp( (1 -\epsilon)^2 \mu/2 ).
\]
 It is straightforward to see that  $\exp((1 -\epsilon)^2 \mu/2 ) <
 n^{-2} $ provided that $\mu > 4 \log n/(1-\epsilon)^2$,
%In conclusion,  
% given that $\epsilon$ can be made  arbitrarily small,
% as long as  $\mu > 4 \log n$  
which corresponds to
 $\alpha_k>\left (\frac{8 \bar{w}
\log n}{Cs^2(1-\epsilon)^2}\right)^{\frac{1}{3-\beta}}$. 

Then, we can claim that $ \P\left( \sum_{[l_1,l_2]\in \Mc_{k-1}}  \ind_{i,l} \leq \rho_k\right) < n^{-2}$ 
provided that for some $\epsilon>0$ jointly
 $\alpha_k> \alpha_k^*=\left (\frac{8 \bar{w}
\log
   n}{Cs^2(1-\epsilon)^2}\right)^{\frac{1}{3-\beta}}$ and
 $\rho_k<(1-\delta)\mu=\epsilon \mu$. The last condition can be
 reformulated in terms of $n$ as\footnote{The second term in the right
 hand side of the inequality can be easily obtained by upper bounding
 $\alpha_k$ with $n^\gamma$.}:
 $n> n_1=\max\left \{ \exp \left [ \left(\frac{8
   \bar{w}
}{Cs^2}\right )^{2-\beta} \epsilon^{\beta-3}\right ],
 \left (\frac{2 \bar{w}
}{Cs^2\epsilon} \right )^{\frac{2}{1-2\gamma}} \right
 \}$. 
% whenever  $\alpha_k^*<\alpha_k<n^{\gamma}$.

At last, jointly considering all pairs in $\Mc_{k}$, 
the probability that  $\sum_{[l_1,l_2]\in \Mc_{k-1}}  \ind_{i,l}  \leq
\rho_k$ for some $[i_1,i_2]\in \Mc_{k}$,
is:
\begin{multline}
 \P\left( \exists [i_1,i_2]\in \Mc_{k} | \sum_{[l_1,l_2]\in \Mc_{k-1}}
   \ind_{i,l}  \leq \rho_k\right) \\ 
\le \sum_{[i_1,i_2]\in \Mc_{k}}  \P\left(\sum_{[l_1,l_2]\in \Mc_{k-1}}  \ind_{i,l} \leq \rho_k\right) <
n n^{-2}=n^{-1}
\end{multline}
provided that jointly $n>n_1$
and  $\alpha_k^*<\alpha_k<n^{\gamma}$, as immediate consequence of probability sub-additivity.
\endproof

%%% PEZZO SOTTO DA REINTRODURRE SE SLICE IMPERFETTE -- begin 
%\begin{comment}
A stronger  statement than the previous one is proved below. 
\begin{corollario} \label{lemma-prop-thorugh-strips-imp}
Consider the good pairs   $[i_1,i_2]\in \Mc_{k}$, with  vertex weight $w_i\in
[\alpha_{k+1}, \alpha_{k}]$. Also, consider the subset $\Mc_{k-1}^*
\subseteq \Mc_{k-1}$, such that 
$ \frac{|\Mc^*_{k-1}|}{|\Mc_{k-1}|}> \eta$, for some $\eta>0$.  Given a generic pair
$[i_1,i_2]\in\Mc_{k}$, for any $\epsilon>0$, 
  with probability  greater than $1- n^{-2}$, 
the number of its  neighboring  good pairs in  $\Mc_{k-1}^*$ 
 is greater than  $\rho_k= \max(4,
 \frac{(\alpha_k)^{4-\beta}}{\sqrt{n}}) $. This result holds 
as long as $ \left (\frac{8 \bar{w}
\log n}{\eta Cs^2(1-\epsilon)^2}\right)^{\frac{1}{3-\beta}}=\alpha_k^*<\alpha_k<n^{\gamma}$ 
 (with $1/4<\gamma<1/2$), and 
$n>n_1/\eta$
Furthermore, the above property 
holds uniformly over the  good pairs in $\Mc_{k}$ with a probability  greater than $1- n^{-1}$, 
under the same conditions as before on $\alpha_k$ and $n$.
\end{corollario}
\proof The proof  follows exactly the lines as the  proof of Theorem \ref{lemma-prop-thorugh-strips}, by replacing $N_{k-1}$ with $\eta N_{k-1}$.
\endproof 
%\end{comment}
%%% DA REINTRODURRE PEZZO SOPRA SE SLICE IMPERFETTE -- end 

Similarly, the theorem below proves that the probability that a bad pair has
a number of neighboring good pairs greater than, or equal to, a given threshold
tends to zero.  
\begin{teorema} \label{lemma-prop-thorugh-strips2}
Consider the  bad pairs   $[i_1,j_2]$, with  vertex weight
$w_i, w_j<\alpha_k$, being
$\alpha_k<n^{\gamma}$   ($\gamma<1/2$). 
Uniformly  over such  pairs  $[i_1,j_2]$,  for any
 $n> n_2=\max\left
  \{\left (\frac{272Cs^4}{\bar{w}^2}\right )^{\frac{2(4-\beta)}{3-\beta}},\left (\frac{36
     Cs^4}{\bar{w}^2}\right )^{\frac{2}{1-2\gamma}} \right \}$, 
 with a probability  greater
than $1- n^{-1}$,
the number of their neighboring good  pairs, $[l_1,l_2] \in \Mc_k$, is 
smaller  than  $\rho_k= \max \left(4, \frac{
  (\alpha_k)^{4-\beta}}{\sqrt{n}} \right)$. 
\end{teorema}
The proof follows the same lines as the proof of
Theorem \ref{lemma-prop-thorugh-strips}; thus, it is omitted for
sake of brevity.
We only remark that now the average number of good  pairs  in $\Mc_k$, which are 
 neighbors of a bad pair  $[i_1,j_2]$, is  $\mu=\Theta\left( \frac{s^2(\alpha_{k})^{5-\beta}}{n}\right) = O(\frac{\alpha_k}{\sqrt{n}} \rho_k)$
with $\frac{\alpha_k}{\sqrt{n}}< n^{\gamma-1/2}$.

Theorems \ref{lemma-prop-thorugh-strips} and 
%Corollary \ref{lemma-prop-thorugh-strips-imp} 
and \ref{lemma-prop-thorugh-strips2} provide the basic tools to show that
the DDM algorithm can   match all good pairs in slices 
 $\Mc_k$ for $k\ge2$, with $\alpha_k >\alpha_k^*=(\frac{8 \bar{w}
\log
   n}{Cs^2(1-\epsilon)^2})^\frac{1}{3-\beta}$.  That is, the good pair matching successfully percolates from one slice
 to the next  till we reach $\alpha_k^*$, without requiring a
 ``local'' seed set in  $\Mc_k$. 
Thus, our algorithm  evolves through stages.
At stage $k+1$, the DDM algorithm fixes $r_{k+1}=\rho_{k}=\max(4, \frac{ \alpha_{k}^{4-\beta}}{\sqrt{n}})$
and  matches  all the previously unmatched pairs of vertices,  
with weight smaller than $\alpha_{k}$, that have at least  $r_{k+1}$ neighbors among the already matched pairs in $\Mc_{k}$.
Observe that the validity of the whole recursion through $k$ is
guaranteed again  by sub-additivity of probability.  I.e., 
given that the number of stages  is by construction upper bounded by
$\frac{\log n}{2}$, for  
$ n >\max (n_1,n_2)$:
\begin{multline}
 \P\left( \exists k | \text{either  not all good  pairs  in $\Mc_k$ are matched } \right. \\
 \left. \text{ or  some bad pair is matched} \right) \le n^{-1} \log n \,. 
\end{multline}

Next,  consider slices  $\Mc_h$ such that  $\alpha_h \le \alpha_k^*$.
 The same algorithm with $r_h=4$ can be  applied,
 however  only a weaker  form of  percolation occurs in this case. 
%as shown by the theorem below.
\begin{teorema} \label{lemma-prop-thorugh-strips3}
Consider the good pairs   $[i_1,i_2]\in \Mc_{h}$, with  vertex weight $w_i\in
[\alpha_{h+1}, \alpha_{h}]$.  
Also, assume that, for some $\eta>0$, at least  a fraction $\eta$   of  neighboring good  pairs, $[l_1,l_2] \in \Mc_{h-1}$,   
%(i.e., pairs whose vertices have  weight   $w_j\in [\alpha_{h-1},\alpha_{h}]$)   
have been previously identified. 
Then, for any $0<\epsilon<1$,   at least a fraction  $(1- \epsilon)$ of  
  pairs  $[i_1,i_2]\in\Mc_{h}$  have a  number of  neighbors among  the  identified pairs $[l_1,l_2] \in \Mc_{h-1}$ greater than 4   w.h.p.,
 as long as $\alpha_{h} \to \infty $. 
\end{teorema}
\proof We exploit again the indicator function    $\ind_{i,l}$  and 
repeat the same  arguments as in the proof of Theorem
\ref{lemma-prop-thorugh-strips}. Then,  given any $0< \eta<1$, we  
 define $\mu=\eta N_{h-1} p_{\min}s^2= \eta C n s^2 \alpha_{h}^{1-\beta}\frac{\alpha_{h}^2}{2n \bar{w}
}= \Theta\left( s^2(\alpha_{k})^{3-\beta}\right) $. 
Since  $ 4 \ll \mu $, 
 we have:
\begin{multline}
 \P\left( \sum_{l\in \Mc_{h-1}, l \text{ identified}}  \ind_{i,l} \le 4 \right)<\\ 
\P(\text{Bi}(\eta N_{h-1}, p_{\min}) \le 4) \le \exp(- \delta^2 \mu/2 )
\end{multline}
with $\delta= \frac{\mu-4}{\mu}$  and as long as $\alpha_h\gg 1$.

Next, let us denote by  $Y_h$ the random variable indicating the
number of vertices in $\Mc_{h}$   
that have at least  $4$  neighbors  among the vertices in $\Mc_{h-1}$
that have been previously  identified.  Then, the above result implies
that: 
$\mathbb{E} [Y_{h}]\ge (1- \exp(- \delta^2 \mu/2 )) N_{h}= N_{h}-
o(N_{h})$. 
Thus, for a sufficiently large $n$  such that $\exp(- \delta^2 \mu/2 )<\epsilon/2$, 
(i.e., $ \mu> \max\left (8,-4\log \frac{\epsilon}{2} \right )$ and
$\mathbb{E} [Y_{h}]> (1-\epsilon/2) N_{h}$),  recalling that $0<\epsilon<1$,   we have:
\begin{equation}
\P(Y_{h}\le (1-\epsilon)N_{h}) <
e^{[-\epsilon^2(1-\frac{\epsilon}{2})\frac{N_{h}}{8}]}\to 0\quad \text
{as $\alpha_{h}\to \infty$} .\nonumber 
\end{equation} \endproof
%\begin{multline}
%\P(Y_{h}\le (1-\epsilon)N_{h})\\
% <  \exp\Big(-\epsilon^2\Big(1-\frac{\epsilon}{2}\Big)
% \frac{N_{h}}{8}\Big) \to 0  \qquad \text {as $\alpha_{h}\to \infty$} .
%\end{multline}\endproof

Furthermore,  consider slices in the interval $h\in[h_{\min},
h_{\max}]$, where $h_{\min}$  has been chosen  so as to guarantee 
$\alpha_{h_{\min}}\ge(\frac{8 \bar{w}
\log
  n}{Cs^2(1-\epsilon)^2})^\frac{1}{3-\beta}$,  
 while  $h_{\max}$ is such that  that $\alpha_{h_{\max}} \to \infty$. 
Then, a sufficiently large $n_3$ can be found such that  uniformly
on $h\in[h_{\min}, h_{\max}]$ we have  $ \mu_h> \max\left (8,-4\log
  \frac{\epsilon}{2} \right )$ 
 (i.e., $\exp(- \delta^2 \mu_h/2 )<\epsilon/2$). This is because, by
 construction,  for every $n$, $\mu_h$ is  decreasing with $h$. Thus,
 if for a given $n$ the expression     
$\mu_{h_{\max}}>    \max\left (8,-4\log \frac{\epsilon}{2} \right )$  
 holds, the relationship is  automatically satisfied for any $ h<h_{\max}$.
Now,  for  $n\ge n_3$,   by sub-additivity   of probability  
we can bound the probability that the DDM algorithm at some stage fails  to 
 identify  at least  a fraction  $1 -\epsilon$  of good
 pairs. Specifically, the bound is given by: 
 $\sum_{h_{\min}}^{h_{\max}} \exp\Big(-\epsilon^2\Big(1-\frac{\epsilon}{2}\Big) N_{h}/8\Big)=  \sum_{h_{\min}}^{h_{\max}}
\exp\Big(-\epsilon^2\Big(1-\frac{\epsilon}{2}\Big) N_{h_{\min}} 2^{(h- h_{\min})(\beta-1)}/8\Big)=
\Theta ( \exp(-\epsilon^2(1-\epsilon/2)N_{h_{\min}+1}/8  )) \to 0 $.  
 We conclude that, for any $\epsilon>0$,  we can iteratively identify
  at least  a fraction  $1 -\epsilon$  of good pairs jointly  in all  slices w.h.p., 
as long as  for  each slice $h$ the assumptions of Theorem \ref{lemma-prop-thorugh-strips3} are satisfied for some $\eta>0$.

%%% PEZZO SOTTO DA REINTRODURRE SE SLICE IMPERFETTE -- begin 
%\begin{comment} 
\begin{teorema} \label{lemma-prop-thorugh-strips2a}
Consider  bad pairs   $[i_1,l_2]$, with  vertex weight
$w_i<2\alpha^*_{k} $ and $w_l<2\alpha^*_k$, (with $\alpha^*_k$ defined as before).
Uniformly  over such  pairs  $[i_1,l_2]$,  for any  sufficiently large   $n$  
%{\bf la soglia esatta non dovrebbe esssere necessaria perche'
%non lo usiamo solo una  volta e non tante volte in catena}
 with a probability  greater than $1- n^{-1}$,
the number of their  neighboring good pairs $[j_1,j_2]$,  with weight $w_j<\alpha^*_k$
  is  smaller  than  $\rho_k= 4$. 
\end{teorema}
\proof  The proof follows exactly the same lines as the proof of
Theorem \ref{lemma-prop-thorugh-strips2} and, thus, it is omitted for
sake of brevity.
Note, however, that now the average number of good  pairs  whose  vertex  weight is  not greater than  $2\alpha_k^*$, which are 
 neighbors of a bad pair  $[i_1,l_2]$ (with $w_i<2\alpha^*_{k} $ and $w_l<2\alpha^*_k$), is upper bounded by 
 $\mu=n 2 (\frac{s^2  2\alpha_k^*}{n \bar{w}
 })^2= o(\frac{\log n}{n}
 )$. 
Thus,  by i) bounding the actual number of neighbors  of  $[i_1,l_2]$
with a binomial distributed r.v., ii)    then applying the Chernoff bound to such variable, and iii)  exploiting sub-additivity  of probability, we get the assert.
\endproof
%\end{comment}
%%% PEZZO SOPRA DA REINTRODURRE SE SLICE IMPERFETTE -- end

At last, we consider  slices $\Mc_q$ such that
$\alpha_q=\Theta(1)$. The following result holds. 
\begin{teorema} 
\label{lemma-prop-thorugh-strips4}
Consider the good pairs   $[i_1,i_2]\in \Mc_{q}$, with  vertex weight $w_i\in
[\alpha_{q+1}, \alpha_{q}]$.  
 A finite fraction  $f(\alpha_{q})$ ($0<f(\alpha_{q})<1$) of such
 pairs 
 have a number of neighbors among  the  identified pairs $[l_1,l_2]
 \in \Mc_{q-1}$  greater than 4,   with a probability  at least  $1-
 n^{-1}$. This result holds 
provided that at least  a fraction $f(\alpha_{q-1})\ge f(\alpha_{q}) $ of   neighboring good pairs $[l_1,l_2] \in \Mc_{q-1}$ 
(i.e., pairs whose vertices have  weight   $w_j\in [\alpha_{q},\alpha_{q-1}]$)   
have been previously identified.
The above  property holds    for   properly selected values of
$f(\alpha_{q})$, 
%= f(\alpha_{q-1})=\frac{99}{100}(1-\exp(72/16))$  
whenever  $\alpha_{q}>(\frac{32\bar{w}
}{Cs^2 f(\alpha_{q})})^{\frac{1}{3-\beta}}$
and  $n>\frac{2\alpha_q^{\beta-1}}{10^4 Cs^2 f(\alpha_{q})}$. 
\end{teorema}
\proof Define $Y_{q}$  as  in   the proof of Theorem
\ref{lemma-prop-thorugh-strips3}.  
If $\EE[Y_{q}]>  (1+\epsilon) f(\alpha_{q}) N_{q} $,  for some $\epsilon>0$, we can claim:  
\begin{equation}
\P(Y_{h}\le f(\alpha_{q})N_{q}) <  \exp\Big(-\epsilon^2\EE[Y_{q}]/2 \Big) < n^{-1}   
\end{equation}
as long as $n>\left (\frac{4 E[w]}{\epsilon^2Cs^2
    f(\alpha_{q-1})}  \right )^2$.
Now, $\EE[Y_{q}]>N_{q}(1- \exp(- \delta^2 f(\alpha_{q})\mu_q/2 ))$  with $\mu_q\ge  Cs^2 \alpha_{q}^{1-\beta}\frac{\alpha_{q}^2}{2 \bar{w}
} $  
 and  $\delta= \frac{f(\alpha_{q})\mu_q-4}{f(\alpha_{q})\mu_q}$.  
Thus, to enforce  $\EE[Y_{q}]>  (1+\epsilon) f(\alpha_{q}) N_{q} $,   we  impose
$ N_{q}(1- \exp(- \delta^2f(\alpha_{q}) \mu_q/2 ))\ge  (1+\epsilon) f(\alpha_{q}) N_{q}$, i.e., $1- \exp(- \delta^2f(\alpha_{q} \mu_q/2 ))\ge  (1+\epsilon) f(\alpha_{q}) $,
from which we can derive  the minimal value of  $\mu_q$ and the
maximal  $f(\alpha_{q})$ for which the previous inequality holds. \endproof
%For example, assuming  $f(\alpha_{q}) \mu_q \ge 16$,   we have $\delta>3/4$ and thus 
%$ (1- \exp(- 9/2))\ge (1+\epsilon) f(\alpha_{q})$,  i.e., $
%f(\alpha_{q})\le (1- \exp(- 9/2))/(1+\epsilon)$ and $ \mu_q \ge
%16/f(\alpha_{q})$. 
%The assert follows by  fixing $\epsilon=1/100.$
%\endproof
As before,  the joint application  of  Theorem  \ref{lemma-prop-thorugh-strips4}    to all slices $\Mc_{q-1}$ with   $\alpha_{q}>(\frac{32\bar{w}
}{Cs^2 f(\alpha_{q})})^{\frac{1}{3-\beta}}$
permits concluding that at  least a fraction of good pairs in each slice $\Mc_{q-1}$
is matched w.h.p while no bad pairs are matched (again thanks to Theorem~\ref{lemma-prop-thorugh-strips2}).
In conclusion, a  fraction of vertices $\Theta(n)$  is successfully identified by our algorithm.  

As last,  the DDM algorithm considers pairs in  slice  $\Mc_0$. 
Theorem \ref{lemma-prop-thorugh-strips5} (whose proof is omitted for brevity) guarantees that all (and only) good pairs in $\Mc_0$ are  matched by our algorithm. 
\begin{teorema} \label{lemma-prop-thorugh-strips5}
Consider a generic pair $[i_1,i_2]\in\Mc_{0}$  with $w_i> n^\gamma/2$, 
and a slice  $\Mc_k$ such that $\alpha_k \le \log^2 n $. 
For a sufficiently large $n$,    with probability  greater than $1- n^{-1}$, 
the number of   good  pairs $[l_1,l_2] \in \Mc_k^*$ 
(with $\Mc_k^*\subseteq \Mc_k$ and  $\frac{|\Mc_k^*|}{|\Mc_k|}>\eta>0$)
 that are  neighbors of   $[i_1,i_2]$
 is greater than  $\rho_0= n^{\gamma/2}$. 
Also,  for  sufficiently large $n$,   with probability  greater than $1- n^{-2}$, 
the number of   neighboring good pairs $[l_1,l_2] \in \Mc_k$  of bad pair  $[i_1,j_2]\in \Mc_0 $
 is smaller than  $\rho_0$, 
The above properties  hold uniformly over all  good pairs in  $\Mc_{0}$ w.h.p.
\end{teorema}
\begin{comment}
\proof
The proof proceeds similarly as before, thus we skip the details and
provide a concise outline.
First, observe  that,  given  a good  pair $[i_1,i_2]\in\Mc_{0}$, 
the average number of  its  identified neighbors pairs   in $\Mc_k$
is, by construction,  
 $\mu\ge N_{k-1}  \frac{n^{\gamma} \alpha_k}{n \E[w]}= \Theta
 (n^\gamma \alpha_k^{2-\beta})=\Omega(n^\gamma log n^{4-2\beta} )$. 
 Similarly, 
given a bad  pair $[i_1,j_2]\in\Mc_{0}$, the average number of
its  identified neighbors pairs   in $\Mc_k$   is, by construction,  
 $\mu\le N_{k-1}  (\frac{ n^{\frac{1}{2}} \alpha_{k}}{n \bar{w}
})^2=
 \Theta(\alpha^{3-\beta})=O(\log n^{6- 2\beta})$.
Thus in both cases,  first bounding from below/above  the  random
variable that expresses the actual number of neighbors  for the considered pair 
and then applying the Chernoff bound, we get the assert.
At last, the fact that  the above property  holds uniformly with
respect to all pairs  descends from  the subadditivity property of probability.
\endproof 
\end{comment}

\subsection{Uniformly distributed seeds\label{subsec:uniform}}
Up to know we have assumed that all the initial seeds in $\Ac_0$
belongs to $\Mc_1$. 
Now, we show that the DDM algorithm can 
 properly percolate when seeds are uniformly  distributed over the slices.
Note that, although the uniform one is the most relevant, our results hold for any arbitrary distribution of the seeds over the graph. 
%We start introducing $\Ac_0$:
%We denote such set of edges by  $\partial \Ac_0$.
We start introducing  the key parameter  that characterizes the ability to start
the bootstrapping percolation  process over $\Mc_1$ (and then over the whole $\Mc$):
\begin{definizione}
 We denote the set of edges between 
the seed set   $\Ac_0$  and the rest of  pairs  $\Mc(\Tc)\setminus \Ac_0$, by  $\partial \Ac_0$.
\end{definizione}
Then we can prove:
\begin{teorema} \label{seeduni}
Whenever the peer set $\Ac_0 $ is chosen  in such a way that: 
\[
|\partial \Ac_0|  \gg  n^{\gamma +\frac{(1- 2 \gamma)r+ \gamma(\beta-1)-1}{r-1}},
\]
our DDM algorithm  percolates identifying $\Theta(n)$ good pairs. 
\end{teorema}

\proof
We proceed as follows. By exploiting the monotonicity property of the percolation
process, 
we can show that a properly dimensioned set of   seeds belonging to slice  $\Mc_k$  $k>1$
is equivalent to a single seed  belonging  to $\Mc_1$. Similar
arguments can be used to show that 
a group of seeds in $\Mc_1$ behaves as a seed in  $\Mc_0$. 
More formally, we consider the evolution of the DDM algorithm
operating on a seed set $\Ac_0$   of pairs  in  $\Mc_1$. Then, we
compare it to the evolution of a modified version of the 
DDM algorithm  operating on a seed set $\Ac_0^*$, which differs from
$\Ac_0$  in that a fraction of seeds in $\Mc_1$ 
 is replaced with a group of seeds,  $S_k$,  in $\Mc_k$.

The modified version of the DDM algorithm handles every 
group of seeds  belonging to $\Mc_k$ as a single seed (i.e., all the
seeds in the same group are 
selected by the algorithm at the same time and  simultaneously included 
in $\Zc$). Also, while proceeding, the two versions of the 
algorithm process exactly the same  sequence of seeds.   
We show that, by properly setting $S_k$, we can  guarantee  that  
the  process of good pairs matching spread faster starting from
$\Ac_0^*$   than from $\Ac_0$. 

Consider a generic good  pair $[i_1,i_2]$ in $\Mc_1$. Note that, by
construction, 
the number of edges between $[i_1,i_2]$ and a given pair $[l_1,l_2] \in \Ac_0$
is either 0 or 1.
The probability that  such edge exists  in $\Mc(\Tc)$  is  upper-bounded by  $p_{1,1}=\frac{w_i \alpha_1}{n \bar{w}
}$.
Instead, the probability that at least an edge exists between
$[i_1,i_2]$ in $\Mc_1$  and the corresponding 
 group of $S_k$ seeds in  $\Mc_k$   
 is lower-bounded  by 
$p_{1,S_k}= 1-(1- \frac{w_i \alpha_{k+1}}{n \bar{w}
})^{S_k}$.
By setting  $S_k> \frac{\alpha_1}{\alpha_{k+1}} + \epsilon$  for any $\epsilon>0$, 
 it can be  easily show that, for sufficiently large $n$, $p_{1,S_k}>p_{1,1}$, i.e., the group of $S_k$  seeds
belonging to  $\Mc_k$  in $\Ac_0^*$   distributes to any good  pair
in $\Mc_1 \setminus \Ac_0$  a number of marks  that upper bounds those distributed by the corresponding 
 seed  in  $\Ac_0$.
This immediately implies that $\GoodB_t(\Ac_0^*)\setminus \Ac_0 \supseteq  \GoodB_t(\Ac_0) \setminus \Ac_0$ for any $t$. 
 Therefore,  at $t_1$ defined as  in Theorem~\ref{propLU},  
 $\GoodB_{t_1}(\Ac_0^*)$ must necessarily include all pairs in
 $\Mc_1\setminus \Ac_0$.  In addition, it is
straightforward to show that every pair  in $\Ac_0\setminus\Ac_0^*$
has at least $r$ neighbors among good  pairs in $\Mc_1\setminus
\Ac_0$ and, thus, it is included  
in $\GoodB_{t_1}(\Ac_0^*)$. 

To conclude the proof, we have to show that
$\BadB_{t_1}(\Ac_0^*)=\emptyset$. This can be done by following the
lines  of Theorem~\ref{propLU}, i.e.,   uniformly upper-bounding the probability of adding marks at
any time $t$ to bad pairs in $\Mc_1$, and, then, repeating the arguments of Corollary \ref{corollary:gross1}.
Then, iterating the previous argument  for all the slices containing seeds,
we get the assertion.
\endproof 
%distributes a number of marks to pairs in $\Mc_1$  which statistically upper bounds  those distributed  by a single seed  belonging  $\Mc_1$  

From Theorem \ref{seeduni},  it immediately descends that, for any choice of seeds, we can correctly match  $\Theta(n)$
good pairs provided that the size of the seed set is at least of order of  $n^{\frac{1}{2}+\epsilon}$, for an arbitrarily small $\epsilon$.

\section{Experimental validation}\label{sec:validation}
Our results hold asymptotically as the number of nodes tends to infinite,
thus it is difficult to validate them considering networks of finite size.  
Nevertheless, in this section, we show that the dramatic impact of power-law degree
on the performance of graph matching algorithms is evident even on small-scale systems.
Another important goal of this section is to check whether 
Chung-Lu graphs, which only capture effects due
to the (marginal) degree distribution of the nodes, can indeed 
predict the performance achievable in real social networks, 
which possess several other features not accounted for by the simple 
Chung-Lu model.

In our first experiment, we take a publicly available, early snapshot
of Facebook containing friendship data of users \cite{facebook}.  
This graph contains 63,371 nodes, the average node degree is 25.64, and the
power law exponent, estimated using the maximum-likelihood approach
\cite{newmanML}, is 2.9412 (quite large). To understand the impact
of network structure, we proceed as follows: we generate a $G(n,p)$ (Erd\"os-R\'enyi) 
graph with the same average degree as the Facebook snapshot, and a Chung-Lu graph
which, besides the average, reproduces also the power-law exponent of the Facebook snapshot, 
using the simple weight sequence introduced in Sec. \ref{subsec:basic}.
We obtain three graphs, which are used as groundtruth network $\Tc$.
We fix the edge sampling probability to $s = 0.7$. 

We run the PGM algorithm on the $G(n,p)$ graph, and a simplified version of the 
DDM algorithm on both the Chung-Lu and the Facebook graphs, considering either
the case of seeds uniformly distributed, or seeds selected only
among nodes whose degree lies in the interval $[\sqrt{n}/2, \sqrt{n}]$.
I.e., we take $\gamma = 1/2$ for the first slice, even though in theory we should 
take a value slightly smaller than $1/2$. For a more meaningful comparison,
our simplified version of DDM employs a constant threshold $r=4$ for all slices,
the same used in PGM. Results are reported on Fig. \ref{fig:figura1}, in which 
we average the number of matched nodes obtained in 100 different runs\footnote{The three graphs are fixed, but randomness is present in the identity of the 
initial seeds and within the algorithms themselves.}. 

\tgifeps{9}{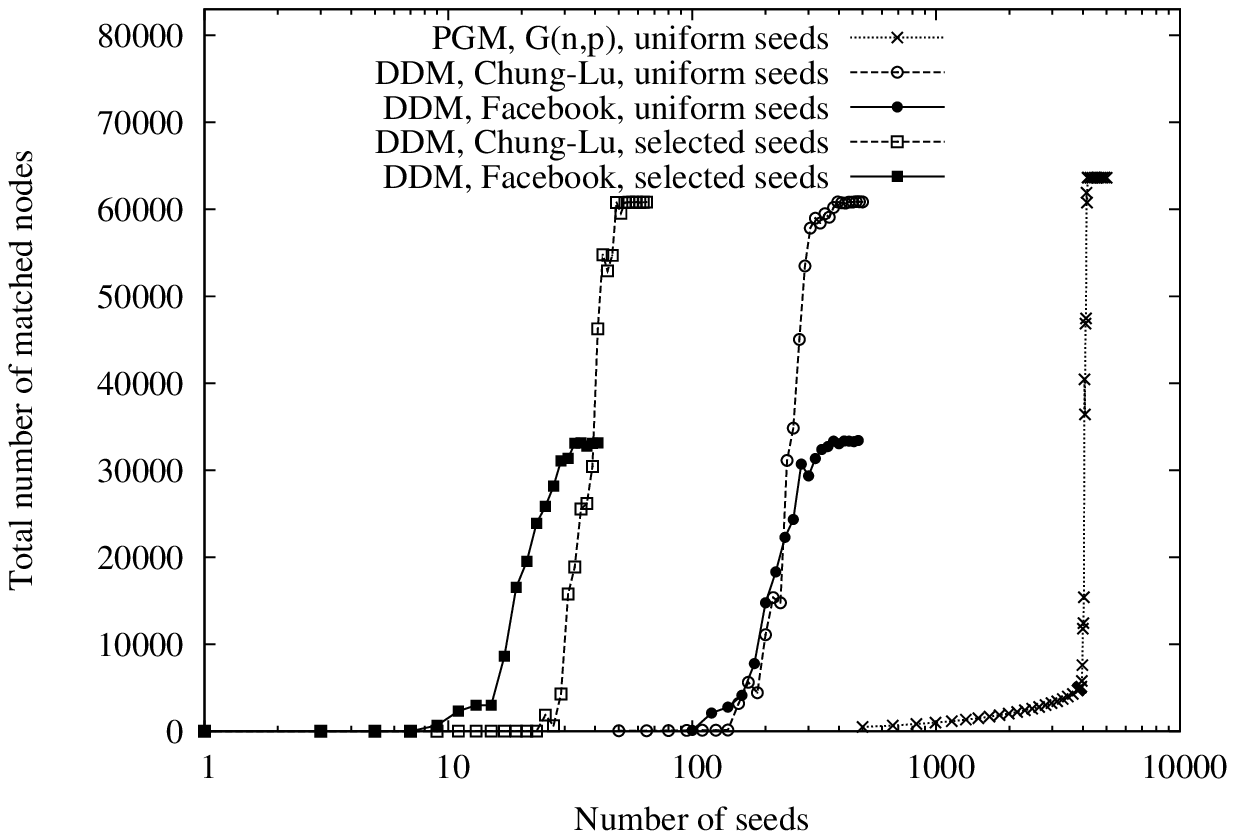}{Total number of matched nodes vs number of seeds,
for different graphs and algorithms, in the case of $s = 0.7$, Facebook social network.}

We clearly see a phase transition effect in all cases, but the position
of the transition changes dramatically (notice the log x scale).
Even a power-law exponent of 2.9 can reduce the threshold 
associated to a $G(n,p)$ graph by more than one order of magnitude,
still considering uniformly distributed seeds. A reduction of another 
order of magnitude is gained by selecting all seeds in the initial
slice of DDM. Very interestingly, the position of the threshold
is more or less the same in the Chung-Lu graph and in the real Facebook snapshot,
meaning that taking into account the power-law exponent alone 
allows us to predict the performance of graph 
matching algorithms in a real social network quite well.

Note that using the Facebook graph the total number of matched nodes
does not go beyond 33K. This is due to the fact that a large fraction of
nodes in this snapshot have degree smaller than 4, hence they cannot
be matched in any case\footnote{This does not occur with the
Chung-Lu graph, in which low-degree nodes are almost not present, 
since we decided to reproduce just the tail behavior (power-law exponent) of the Facebook 
degree distribution.}.
At last, we report some figures for the fraction of 
bad pairs matched by our algorithm in the above experiment (negligible errors
were produced by PGM in the Erd\"os-R\'enyi graph).
We consider only the fraction of bad pairs at the phase
transition point, because here the error is known to be maximum 
\cite{Grossglauser}. We observed about 0.001 (0.0002) fraction of bad pairs
using the Chung-Lu graph, respectively with uniform and selected seeds.
The Facebook snapshot produced slightly more matching errors. However,  
%0.05 and 0.02, respectively, for reasons that we are yet to
%understand. 
we do not consider these errors
really significant, as they could be reduced by a more careful selection 
of threshold $r$, without affecting the scaling-order 
performance gains of our algorithm.  

\tgifeps{9}{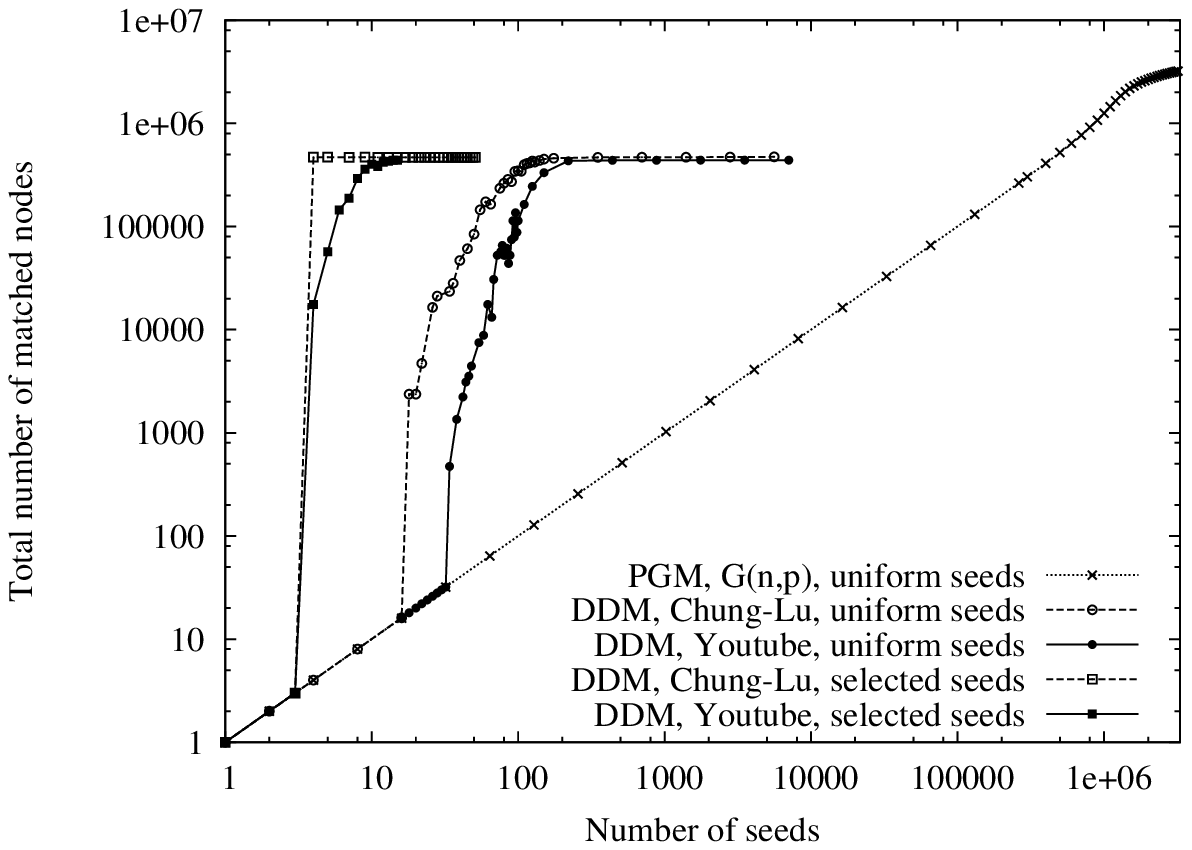}{Total number of matched nodes vs number of seeds,
for different graphs and algorithms, in the case of $s = 0.9$, YouTube social network.}

In our second experiment, we used a social network graph 
representing friendship connections among YouTube users \cite{mislove}.
This graph contains 3,223,589 vertices, the average node degree is 5.81, and the
power law exponent, estimated using the maximum-likelihood approach
\cite{newmanML}, is 2.23. Similarly to what we did in our first experiment,
we generated a $G(n,p)$ graph with the same average degree as the YouTube graph, and a Chung-Lu graph
which, besides the average, reproduces also the power-law exponent of the YouTube.
This time we used an edge sampling probability $s = 0.9$. 
The performance of PGM and DDM algorithms (for uniform and selected seeds)
on the above graphs is reported on log-log scale in Fig. \ref{fig:figura2}, in which 
we averaged the results of 100 runs. 
We observe that a phase transition is barely visible in the case of the 
$G(n,p)$ graph, where the final number of matched nodes is always only slightly larger
than the number of seeds. The percolation phenomenon is instead clearly
visible on both Chung-Lu graph and the real YouTube graph, and, again, 
the position of the threshold is surprisingly similar in these two graphs,
both in the case of uniformly distributed seeds and in the case of selected seeds.   
In the latter case (i.e., selected seeds), we observe that
4 seeds (the minimum number of seeds to trigger a bootstrap percolation
with threshold $r=4$) are essentially enough to identify a large fraction of the nodes
(actually, those nodes having sufficiently large degree to be identified).

\section{Conclusions}
We investigated the problem of user identification in social networks    
represented by  scale-free graphs, by adopting bootstrap percolation and a novel graph slicing technique.
Our major results  show that, for a successful identification, the seed set can be as small as $n^{\epsilon} $ (for any $ {\epsilon>0}$) when
seeds are properly selected, and of the order of  $n^{\frac{1}{2}+\epsilon}$ when they are randomly taken. 
Such findings are confirmed by numerical results obtained with an early Facebook snapshot thus showing that the 
class of scale-free graphs we considered are a good representation of real-world social networks.

%\appendices
\appendix

\begin{comment} 
\section{Proof of Corollary \ref{corollary:gross1} \cite{Grossglauser}\label{app:gross1}}
\begin{multline}
\P \{\BadB_{t_0}\neq \emptyset \} = \P ( \sum_{t=0}^{t_0} \sum_{i,j,i\neq j} \ind_{\{\text{Marks}_{[i_1,j_2]}(t)= r\}} \ge 1)  \\ \le   
\sum_{t=0}^{t_0} \sum_{i,j:i\neq j} \P ( \ind_{\{ \text{Marks}_{[i_1,j_2] }(t)= r\}}=1 ) \\
=\sum_{t=0}^{t_0} \sum_{i,j:i\neq j}  \P \left(\text{Bi}(t,p^2s^2)=r\right)\\
= \sum_{t=0}^{t_0} \sum_{i,j:i\neq j}	\dbinom{t}{r} (ps)^{2r}(1- (ps)^2)^{t-r}\\
 \le  \sum_{t=0}^{n} n^2 (tp^2s^2)^r\le n^{-r \epsilon} .
\end{multline}
\end{comment} 

\section{On the accuracy of graph slicing\label{app:slices}}
In Sec. \ref{sec:problem}, we considered that the pairs graph $\Mc(\Tc)$ is  sliced into
subgraphs $\Mc_k$. The exact procedure would instead
imply that $\Gc_1$ and $\Gc_2$ are sliced according to the observable
degree, and that the pairs graphs  corresponding to the graphs
slices are considered. 
Here, we  show that the effect due to imperfect graph slicing can be made negligible. 
%any two slices obtained from $\Gc_1$ and $\Gc_2$
%as above,  $\gc_{1,k}$ and $\gc_{2,k}$, include 
%the same vertices w.h.p. Thus, considering $\Mc_k$ is equivalent to
%considering $\Mc'_k$.

As the first step, recall that the vertex weight can  be just inferred from the
actual degree, hereinafter  referred to  as estimated weight.  
E.g.,   given a vertex $i_1$ in  $\Gc_1$  with degree $D_1^i$,
the estimated weight associated to it  is 
$\hat{w}^1_i=D_1^i/s$. By  slicing  $\Gc_1$ on the only basis of
such estimated weights, it is clear that each slice may include vertices with different weight than expected.
A similar observation holds for $\Gc_2$.
Then, we are going to show how to build an imperfect slice $\Mc_k'$ with estimated weights in
the range  $[ \alpha_{k+1}, \alpha_{k}]$, such that the following
three 
conditions are satisfied.
1) Only pairs  formed by  vertices  whose actual weight is  in the interval   $[ \alpha_{k+1}, \alpha_{k}]$ are included in  $\Mc_k'$; 
2)  Only a finite fraction of good  pairs of   $\Mc_k$  is not included in $\Mc_k'$; 
3)  The following situation occurs with negligible probability:
 a bad pair $[i_1,j_2]$  is included in the slice while none of the pairs
 $[i_1,i_2]$ and $[j_1,j_2]$ are included.
The third condition ensures that every bad pair in
$\Mc_{k}'$ conflicts with at least one good pair in $\Mc_{k}'$,
thus it cannot be matched by the DDM algorithm 
  as it  (eventually) reaches the threshold. 
%This is  because the  bad pair will
%be blocked by an already matched conflicting good pair.

To let the above three conditions hold, let us build $\Mc_k'$ as
follows. We  partition the  interval $[\alpha_{k+1}, \alpha_{k} ]$,
 into  two sub-intervals.  An interval  $[\alpha_{k+1}(1+\epsilon),
 \alpha_{k}(1-\epsilon)]$,  with $0<\epsilon\le 1/4$, 
is defined as  inner region, while  the remaining range of values
% $[ \alpha_{k+1}, \alpha_{k}]  \setminus[  \alpha_{k+1}(1+\epsilon),
% \alpha_{k}(1-\epsilon)]$, 
is defined  as outer region.
The idea is to include in $\Mc_{k}'$   pairs of vertices whose weights fall in 
either the inner or  the outer region, adding  the  extra constraint 
that only pairs for which at least one vertex falls in the inner
region  are included in   $\Mc_{k}'$.  
This expedient  implies that $[i_1,j_2]$ is 
included in $\Mc_{k}'$ only if  
$i_1$  ($j_2$) falls  in the inner region and $i_2$  ($j_1$) falls in the inner plus outer region.

%\begin{comment}  Nella versione corta al posto di cio' che segue
%Then, by applying standard concentration results, we can easily  show
%that, as long as $\alpha_{k+1}> \frac{65}{\epsilon^2} \log n$ for
%sufficiently large $n$, 
%the above conditions 1), 2) and 3) are satisfied with a probability greater than $1-n^{-1}$.
%\end{comment}
%1)   all vertices  whose weight fall in the inner slice are included in $\Mc_{k'}$ .
%2)   none of the vertices  whose weight fall outside the outer plus inner slice are included in $\Mc_{k'}$ w.h.p.
%3)  the following event: \{There is  a node $i_1$  ($j_2$)    falling  in the inner region, whose  `companion'  node    $i_2$  ($j_1$)  falls outside 
%both the inner and outer regions\} has  a  negligible probability as $n$ grows large
% for a properly chosen $\epsilon$.
%
%\begin{comment}
Since the proofs of points 1) and 2) are trivial we omit them and we limit ourselves  to show 3).
We proceed as follows. 
For a generic  vertex $i$ with weight $w_i\geq \alpha_{k+1}$, we   bound the difference $|\hat{w}^1_i- \hat{w}^2_i |= 1/s|D^i_1-D^i_2|$  
between the estimated weights associated with
  $i_1 \in \Vc_1$ and $i_2 \in \Vc_2$, respectively.
We define as $X^{(1)}_{i,j}$ the indicator function that is equal to 1
if vertex $i_1 \in \Vc_1$  has 
an edge with a generic other vertex $j$ and it is equal to 0
otherwise. Similarly, we define  $X^{(2)}_{i,j}$ for $i_2 \in \Vc_2$.   Also, 
let $D^i_1=\sum_{j\in \Vc_1} X^{(1)}_{i,j}$ and $D_2=\sum_{j\in \Gc_2}
X^{(2)}_{i,j}$. Since $X^{(1)}_{i,j}$  and $X^{(2)}_{i,j}$ are
Ber$(\frac{w_i,w_j}{W})$ random variables, $D^i_1$ and $D^i_2$ are 
identically distributed with equal mean value $\EE[D^i]$.    
Thus, since $\EE[D^i_1]=\EE[D^i_2]=\EE[D^i]$, for any $\eta>0$  the following inequality  holds:
\begin{multline}
\PP(|D^i_1-D^i_2|>2\eta) \\
 \leq \PP(|D^i_1 -\EE[D^i_1]|>\eta) + \PP(|D^i_2 -\EE[D^i_2|>\eta) \\
 = 2 \PP(|D^i_1 -\EE[D^i]|>\eta) .
\end{multline}
By applying Chernoff's bound, we obtain:
\begin{equation}
\PP(|D^i_1 - \EE[D^i]|>\eta) \leq e^\frac{-\eta^2}{2(E[D^i]+\eta/3)}. 
\end{equation}

%\begin{comment}
%The above expressions hold  for any vertex of $\gc_1$. 
%Thus, we can choose  $\eta=\alpha_{k+1}\epsilon/2$ and apply  the
%union bound with respect to vertices so as to get: 
%\begin{multline}
%\PP(\cup_i \{ |D^i_1-D^i_2|>\epsilon \alpha_{k+1} \})\\
%\le \PP(\cup_i \{ |D^i_1 - \EE[D^i]|>\epsilon \alpha_{k+1}/2\}) \\ 
%\le \sum_i 2 e^{\frac{-(\epsilon \alpha_{k+1}/2) ^2}{2(E[D^i]+ \epsilon \alpha_{k+1}/6)}}\
%\le 2n  e^{\frac{-(\epsilon \alpha_{k+1}/2) ^2}{2 (n^{1/2}+ \epsilon \alpha_{k+1}/6)}} \  
%\end{multline}
%with 
%\[
% 2n  e^{\frac{-(\epsilon \alpha_{k+1}/2) ^2}{2 (\alpha_k+ \epsilon \alpha_{k+1}/6)}}\to 0
%\]
%whenever $\alpha_{k+1}\ge $.
%Thus, we can conclude that with high probability no  vertex pair $[i_1,i_2]$ exists in the graph such that  
% $i_1$ $(i_2)$ falls  in the inner region and $i_2$ ($i_i$) falls outside the inner plus  outer region.
%\end{comment}

Restricting for the moment the analysis only to  those vertices  $i$ such that $w_i\le 2\alpha_k$, we can write:
\begin{multline}
\PP(\cup_{i:w_i\le 2 \alpha_k}\{|D^i_1-D^i_2|>\epsilon \alpha_{k+1}\})\\
\le 2\PP(\cup_{i: w_i\le 2 \alpha_k}\{ |D^i_1 - \EE[D^i]|>\epsilon \alpha_{k+1}/2\}) \\ 
\le \sum_{i: w_i\le 2 \alpha_k} 2 e^{\frac{-(\epsilon \alpha_{k+1}/2) ^2}{2(E[D^i]+ \epsilon \alpha_{k+1}/6)}}\
\le 2n  e^{\frac{-(\epsilon \alpha_{k+1}/2) ^2}{2 (2\alpha_k+ \epsilon \alpha_{k+1}/6)}} < n^{-1}  
\end{multline}
whenever $\alpha_{k+1}\geq\frac{65}{\epsilon^2} \log n$.
For what concerns vertices with $w_i> 2\alpha_k$,  again applying the Chernoff bound we get:
\[ 
\PP(\cup_{i: w_i > 2 \alpha_k} \{\hat{w}_1^i < \alpha_k \}\cup \{ \hat{w}_2^i <\alpha_k\})\le 
 2n  e^{-\frac{\alpha_{k}}{4}}\to 0
\]
under the condition that   $\alpha_{k} > 4\log n$.
Thus, we can conclude that with high probability no  vertex pair $[i_1,i_2]$ exists in the graph such that  
 $i_1$ $(i_2)$ falls  in the inner region and $i_2$ ($i_i$) falls outside the inner plus  outer region whenever
 $\alpha_{k+1}\ge \frac{65}{\epsilon^2} \log n$.

In conclusion  previous algorithm can be applied to identify an imperfect slice, which 
w.h.p. satisfies the assumptions  of Corollary \ref{propLU-1} 
(i.e.,  it comprises exclusively  a finite fraction of  pairs in $\Mc_1$ satisfying 
 the constraint that  no bad  pair $[i_1,j_2] \in \Mc_1'$  if none of the pairs    $[i_1,i_2]$ and $[j_1,j_2]$ are in $\Mc_1'$).

%\end{comment}

Now,   Theorem \ref{propLU} can be extended to show that 
our DDM algorithm  correctly percolates within 
slice $\Mc_1'$  provided that $\Mc_1'$  satisfies conditions 1), 2) and 3).
%(a detailed proof  is reported in \cite{techrep}). 
 Similarly,  we  show  
that 
 the above described cascading process  through slices of good pairs matching takes place when slices are  imperfect.
The important condition is that 
%In particular, however in case of imperfect slicing we need  to 
the {\em seed set at every stage} of the algorithm is adjusted
so as to ensure that  conditions 1), 2) and 3) are met. %(see \cite{techrep} for a detailed explanation).

%\begin{comment}
Let us now  summarize the main steps of our modified DDM algorithm to account for the fact that slices are imperfect.
First,  for a suitable $\gamma=1/2 - \epsilon$,  the algorithm  ``extracts''  from $\Mc(\Tc)$
 a core set of  pairs  $\Mc_{1}'$  belonging to slice  $\Mc_1$ and
 satisfying the conditions  1), 2) and 3).
 %of Corollary \ref{propLU-1}.
This is done by applying the algorithm described below and  fixing a small $\epsilon$.
%~\ref{app:slices}  

Second,  we apply the standard PGM algorithm 
%described in Section \ref{subsec:Gnp-pre-res}   
to slice 
$\Mc_{1}'$ in order to successfully identify  all correct pairs within $\Mc_{1}'$.
Slice $\Mc_0'$  is filled with all the pairs  that have not
been placed in $\Mc_{1}'$ and
  for which the expected degree of at least one vertex exceeds  threshold $(\alpha_1 + \alpha_2)/2$. 
This guarantees w.h.p that $\Mc_0\subseteq \Mc_0'$ as well as 
that $\Mc_k\cap \Mc_{0}'=\emptyset$ for any $k>2$.
Third,  we fix $r=\rho_1$ and match all remaining pairs  (i.e., pairs in    $\Mc(\Tc)\setminus (\Mc_{1}'\cup  \Mc_{0}')$)
 that have at least $\rho_1$ neighbors among the matched pairs in $\Mc_{1}'$.  Let us call this set   $\Mc_{2}''$.
By applying  Corollary 
\ref{lemma-prop-thorugh-strips-imp}, which requires that only a finite
fraction of correct pairs in $\Mc_1$  have been matched,
we can  guarantee that every correct pair in $\Mc_2$ 
also belongs to  $\Mc_{2}''$.  
%(as well  as   correct pairs eventually in $\Mc_{1}\setminus (\Mc_{1'}\cup \Mc_{0'}$ are in $\Mc_{2}''$.)  
Furthermore, using Theorem \ref{lemma-prop-thorugh-strips2}, we can
claim that no  bad pair 
falls in $\Mc_{2}''$.  Then, applying the algorithm described below 
%Appendix~\ref{app:slices} 
to matched pairs,  we can ``extract`` a subset $\Mc_2' \subseteq \Mc_{2}''$ satisfying  the following two properties: 
 i) every pair in $\Mc_2'$ belongs to   $\Mc_2$, ii) pairs in
 $\Mc_2'$ are a finite fraction of  all correct pairs in 
 $\Mc_2$.  
As the next step, we set  $r_2=\rho_2$ and match all pairs in    $\Mc(\Tc) \setminus (\Mc_2''\cup \Mc_{1'}\cup  \Mc_{0'})$ 
that have at least  $\rho_2$ neighbors among the matched pairs in $\Mc_{2}'$.
The algorithm is then iterated  for every slice $\Mc_k$,  with $\alpha_{k} > \max(2\frac{65}{\epsilon^2}\log n, \alpha_k^*)$.  
So doing,  % we can  invoke  Corollary \ref{lemma-prop-thorugh-strips-imp} 
 we can  show that every  correct pair in $\Mc_k$   ($k>1$) is  matched (while no bad pairs are
matched thanks to  Theorem \ref{lemma-prop-thorugh-strips2}).

%Observe that the validity of the whole recursion through $k$ also in this case is
%guaranteed   by sub-additivity of probability.  That is, 
%given that the number of stages $k$ is by construction upper bounded by $\frac{1}{2}\log n$, we have:
%\begin{multline}
% \P\left( \exists k | \text{either  not all  correct pairs  in $\Mc_k'$ are matched } \right. \\
% \left. \text{ or  some spurious pair is matched} \right) \le n^{-1} \log n, 
%\end{multline}
%for  $n>\max \left( \exp((\frac{8 \bar{w}}{\eta Cs^2\epsilon^{4-\beta}}\right)^{\frac{1}{3-\beta}}), 
%(\frac{2 \bar{w}}{\eta Cs^2\epsilon})^{\frac{2}{1-2\gamma}}, (\frac{272Cs^4}{\EE^2[w]})^{\frac{2(4-\beta)}{3-\beta}},(\frac{36 Cs^4}{\EE^2[w]})^{\frac{2}{1-2\gamma}}\right)$, 

We now turn our attention to slices $\Mc_h$,  with   
$\alpha_{h}\le \max (2\frac{65}{\epsilon^2} \log n, \alpha_k^*)$. The DDM algorithm operates in the following way.
It fixes the threshold to $r_h=4$ and starts considering  an initial
set of  matched pairs  $I_{h=k^*}= \Mc_{k*}'$,  
where $k*= \arg \max \{\alpha_k >\max(2\frac{65}{\epsilon^2} \log n, \alpha_k^*)\}$ and 
%for which previous step was executed matches all pairs  that have more than 4 neighbors in 
%$I_{h=k^*}= \Mc_{k*}'$; 
we match all pairs that have at least $r_h$  neighbors in $I_{h=k^*}$.
Let $\Mc_{h=k^*+1}''$ denote the set of matched pairs,  by Theorem
\ref{lemma-prop-thorugh-strips3}, 
 $\Mc_{h=k^*+1}''$  contains an arbitrarily large fraction of correct pairs in   $\Mc_{h=k^*+1}$. 
Then,   set $I$ is updated according to the recursion:  $I_{h}=
I_{h-1}\cup \Mc_{h}''$ ,  
and, again, the matching  procedure is iterated  to identify  correct
pairs in the next slice 
for any $h$ such that $\alpha_h\to \infty$.
% The procedure is iterated over  $h$ such that $\alpha_h\to \infty$. 
With arguments similar to those of Theorem~\ref{lemma-prop-thorugh-strips2}, it can be shown that at  no stage of the algorithm any bad pair is matched,
 while  Theorem  \ref{lemma-prop-thorugh-strips3}  guarantees that an arbitrarily large fraction of correct pairs in 
  $\Mc_{h}$ are matched within step $h$ (this because by  construction $I_{h-1}$ contains an arbitrarily large fraction of 
correct pairs in $\Mc_{h_{m-1}}$). 
The same algorithm is then applied to slices $q$  with $q>q_*$,  
in order  to identify in each of such slices at least a fraction
$f(q)$ of correct pairs, according  to Theorem  \ref{lemma-prop-thorugh-strips4}.
At last, we set  $\rho_0= n^{\gamma/2}$ and match   pairs in $\Mc_0'$
that have at least  $\rho_0$ neighbors in one of the 
slices $\Mc_k'$,  for $k$ satisfying $\alpha_{k}= \max (13000 \log n, \alpha_k^*)  <\alpha_k<\log^2 n$.
A strengthened version of Theorem~\ref{lemma-prop-thorugh-strips5}
again guarantees that every  correct pair in $\Mc_0'$ is matched, 
while no bad pairs are matched.
%\end{comment}

\end{sloppypar}
\end{document}